\documentclass[
  journal=aas,
  manuscript=research-paper-preprint, 
  year=2020,
  volume=37,
]{cup-journal}

\usepackage{xpatch}
\makeatletter
\def\cup@journal@name{}
\xpatchcmd{\@maketitle}{(\cup@year), {\volumefont\cup@vol}, \thepage--\pageref{LastPage}}{}{}{}
\xpatchcmd{\@maketitle}{\hfill\includegraphics[width=26mm]{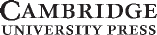}}{}{}{}
\def\ps@plain{%
  \renewcommand{\@oddhead}{\hfill\thepage}%
  \renewcommand{\@evenhead}{\hfill\thepage}%
  \renewcommand{\@evenfoot}{}%
  \renewcommand{\@oddfoot}{\@evenfoot}%
}
\makeatother

\usepackage{amsmath}
\usepackage[nopatch]{microtype}
\usepackage{booktabs}

\usepackage{graphicx}
\graphicspath{{figures/}}
\usepackage{multicol,multirow}
\usepackage{amsmath,amssymb,amsfonts}
\usepackage{mathrsfs}
\usepackage{amsthm}
\usepackage{rotating}
\usepackage{appendix}
\usepackage{ifpdf}
\usepackage[T1]{fontenc}
\usepackage{textcomp}
\usepackage{xcolor}
\usepackage{lipsum}
\usepackage[colorlinks,allcolors=blue]{hyperref}

\theoremstyle{definition}

\numberwithin{equation}{section}

\usepackage[utf8]{inputenc}
\usepackage{float}
\usepackage{hyperref}
\usepackage{graphicx}
\usepackage{makecell}

\usepackage{geometry}
\usepackage{amssymb}
\usepackage{caption}
\usepackage{subcaption}
\usepackage{cmll}
\usepackage{bbm}
\usepackage{array, booktabs, boldline} 
\usepackage{amsmath}
\usepackage{changepage}

\usepackage[textwidth=1.8cm, textsize=tiny]{todonotes}

\title{Contributions of geolocated weather and building related data for insurance assessment of flood risks}

\author{Mulah Moriah}
\affiliation{Laboratoire de Math\'ematiques de Bretagne Atlantique, UBO, Brest, France}
\alsoaffiliation{Euro-Institut d'Actuariat Jean Dieudonné (EURIA), UBO, Brest, France}
\email[Mulah Moriah]{mulah.moriah@etudiant.univ-brest.fr}

\author{Franck Vermet}
\affiliation{Laboratoire de Math\'ematiques de Bretagne Atlantique, UBO, Brest, France}
\alsoaffiliation{Euro-Institut d'Actuariat Jean Dieudonné (EURIA), UBO, Brest, France}

\author{Pierre Ailliot}
\affiliation{Laboratoire de Math\'ematiques de Bretagne Atlantique, UBO, Brest, France}
\alsoaffiliation{Euro-Institut d'Actuariat Jean Dieudonné (EURIA), UBO, Brest, France}

\author{Philippe Naveau}
\affiliation{Laboratoire des Sciences du Climat et de l’Environnement, PSL \& U Paris-Saclay, Gif-sur-Yvette, France }

\author{Juliette Legrand}
\affiliation{Laboratoire de Math\'ematiques de Bretagne Atlantique, UBO, Brest, France}
\alsoaffiliation{Euro-Institut d'Actuariat Jean Dieudonné (EURIA), UBO, Brest, France}


\begin{document}

\begin{abstract}
Floods rank among the costliest natural hazards, causing over USD 100 billion in insured losses from 2013 to 2023. In France, persistent deficits in the natural catastrophe scheme highlight the need for accurate, building-scale flood risk evaluation. Although insurers commonly rely on frequency–severity models supported by hazard maps and regional climate indicators, previous studies show that such large-scale variables explain only a limited share of individual flood loss variability. This study assesses the marginal contribution of multiple georeferenced data layers to modelling flood claim occurrence and severity in a large French home-insurance portfolio. Starting from a baseline model using standard underwriting information, we sequentially introduce climate-expert variables, extreme rainfall indicators, and fine-scale geolocated building and environmental attributes. The analysis targets a practical use case in which insurers cannot deploy full hydrological or hydraulic catastrophe models due to budgetary, licencing, or operational constraints. Results show that rainfall-based indicators, particularly a newly constructed metric capturing intense local precipitation, substantially improve claim modelling, while building and environmental variables further refine occurrence prediction. Overall, findings demonstrate how high-resolution geolocated data enhance exposure and vulnerability assessment, complement official flood maps, and provide insurers with an operational framework for refining flood risk evaluation and pricing.
\end{abstract}

\section{Introduction}

\subsection{Flood risk and the role of insurance}
Insurance is a key instrument for limiting the financial and social costs of climate-related hazards. By pooling risks and spreading losses across policyholders, insurers support households and public authorities in recovering from damaging events. In France, standard home insurance is supplemented by the Catastrophes Naturelles (CatNat) scheme, created in 1982 (\cite{grislain2010regime}), which offers nationwide protection against major natural hazards through a compulsory surcharge and a public–private reinsurance arrangement. This framework has recently faced mounting strain. Climate change is increasing both the frequency and intensity of heavy rainfall, while urban growth and land-use shifts are amplifying exposure and vulnerability. Consequently, the CatNat scheme has posted underwriting losses for eight years in a row, underscoring the need for better methods to quantify, forecast, and price flood risk (see \cite{assureurs2023eventnat}). Floods are the most expensive natural hazard in France, with cumulative CatNat payouts surpassing €$28.8$ billion since 1982 (\cite{assureurs2021impact}). Flood dynamics vary widely, from rapid runoff and flash floods to slower river flooding, and are strongly shaped by local topography, soil conditions, and the built environment, making building-level loss estimation particularly difficult.

\subsection{Existing approaches and their limitations}

Flood modelling traditionally relies on hydrological and hydraulic approaches that simulate runoff processes, river flow dynamics, water depths, flow velocities, and inundation extents. These physical based models form the scientific foundation of flood hazard mapping and are extensively applied in catastrophe modelling frameworks. They usually require high-resolution topographic data, information on hydrological networks and land use, as well as specialised modelling skills. The outputs are then linked with damage functions to estimate potential economic losses, often using aggregated data originating from insurance portfolios or building expertise. Large-scale applications of flood modelling and loss estimation include studies carried out by CCR (Caisse Centrale de Réassurance), the French public reinsurer, and by France Assureurs, the French insurance federation. (see, for example, \cite{assureurs2021impact}).

Insurance claim data, including flood occurrence records and underwriting information, are increasingly recognised as a valuable resource for assessing hazard impacts. Such data have been used to validate hazard maps and rainfall indicators, particularly at aggregated spatial scales such as municipalities or regions (e.g. \cite{zhou2013verification, spekkers2014decision}). While effective for large-scale risk assessment, these approaches are less suited to explain the strong heterogeneity in losses observed at the portfolio or individual-building level.

In practice, most insurers do not have the financial, technical, or operational capacity to deploy detailed hydrological or hydraulic models for routine pricing or portfolio risk management. This limitation is compounded by the fact that insurance portfolios are spatially concentrated rather than uniformly distributed, making local exposure effects especially important. As a result, insurers must assess flood risk at the scale of individual policies, while taking into account the characteristics of the building and the highly localised loss drivers. Consequently, flood risk pricing typically relies on statistical frequency–severity models built on underwriting data, sometimes complemented by hazard maps or broad climate indicators. However, empirical evidence shows that such large-scale variables explain only a limited share of loss variability at the building level (\cite{zhou2013verification, spekkers_2011, merz2013multi}). Local factors such as terrain slope, soil impermeability, proximity to watercourses, surrounding land use, and spatial concentration of intense rainfall play a dominant role in flood damage but are poorly captured by municipal or regional indicators. Consistent with this, \cite{merz_2004} show that water-related variables alone account for only a small fraction of the variability of observed damage, highlighting the need for richer, fine-scale contextual information.

Within the flood-risk framework used by the IPCC (Intergovernmental Panel on Climate Change) and UNDRR (United Nations International Strategy for Disaster Reduction), risk arises from the interplay of hazard, exposure, and vulnerability. In insurance contexts, exposure is typically well approximated by geocoded policy locations. Vulnerability is partly represented through underwriting variables, yet these often lack the spatial and structural detail required to capture building-level features and surrounding environmental conditions. Among the three components, the local dynamics of rainfall intensity and runoff is generally the most poorly represented in standard insurance models.

This study concentrates on modelling flood-claim occurrence and severity within a traditional insurance pricing framework, with the dual objective of estimating flood risk for rating purposes and assessing the risk of contracts already held in the portfolio. The analysis focusses on a practical use case in which insurers cannot deploy full hydrological or hydraulic catastrophe models due to budgetary, licencing, or operational constraints. We seek to improve the conventional insurer flood-risk assessment by systematically incorporating geolocalized and meteorological contextual data, while maintaining the interpretability and day-to-day usefulness of classical frequency–severity statistical models. To our knowledge, no earlier work has simultaneously integrated policy-level insurer records, detailed building and environmental characteristics, and rainfall-based variables to predict flood-related building losses in a statistical approach. Existing studies have generally used aggregated claims data or limited information on individual structures. For instance, \cite{torgersen2017evaluating} included detailed data on building surroundings but omitted weather variables and specific building attributes. By filling this gap, our study provides a more fine-grained view of building-level flood risk and offers actionable insights for insurers and researchers aiming to enhance flood risk modelling and mitigation efforts.

\subsection{Agenda and objectives}

The aim of this study is to evaluate how gradually incorporating consistent, geolocated contextual information can enhance traditional building-level flood-risk models used by insurers. We take the viewpoint of an insurer that depends on statistical frequency–severity models for pricing, monitoring its portfolio, and segmenting risk, in an environment where comprehensive physical hydrological or hydraulic models cannot be routinely applied. For this purpose, we develop a series of nested models. The first model acts as a benchmark and uses only conventional underwriting variables and policy-level data drawn from an insurance portfolio. 

In a second step, we enhance the baseline model by adding hydrological and climatic indicators that are widely used in insurance practice, such as hazard maps, zoning information, and aggregated geographic variables. This layer mirrors the prevailing industry approach to integrating flood hazard data into statistical pricing models. The third modelling layer adds rainfall-based variables derived from gridded meteorological datasets. These variables are designed to capture local hazard dynamics more accurately and to reflect the spatial and temporal variability of precipitation events that static hazard maps cannot represent. Finally, we include detailed geolocated information on building characteristics and their surrounding areas. By incorporating local structural and environmental attributes, this layer aims to explain a larger share of the variability in predicted flood occurrence and associated costs.

At each step, we assess how much the newly added data layer improves the modelling of flood claim occurrence and severity, using performance metrics tailored to insurance contexts. This layered procedure enables us to measure the operational contribution of each information category. The remainder of the paper is structured as follows. Section~\ref{sec:data} describes the insurance portfolio and details the construction of meteorological, building-level, and environmental variables. Section~\ref{sec:modeling} outlines the modelling framework and examines the incremental performance improvements achieved at each stage, as well as their implications. Section~\ref{sec:conclusion} discusses the broader implications of our findings for insurance practice, along with the study’s limitations and avenues for future research.

\section{Data}\label{sec:data}

\begin{figure}[!h]
     \centering
    \includegraphics[width=14cm]{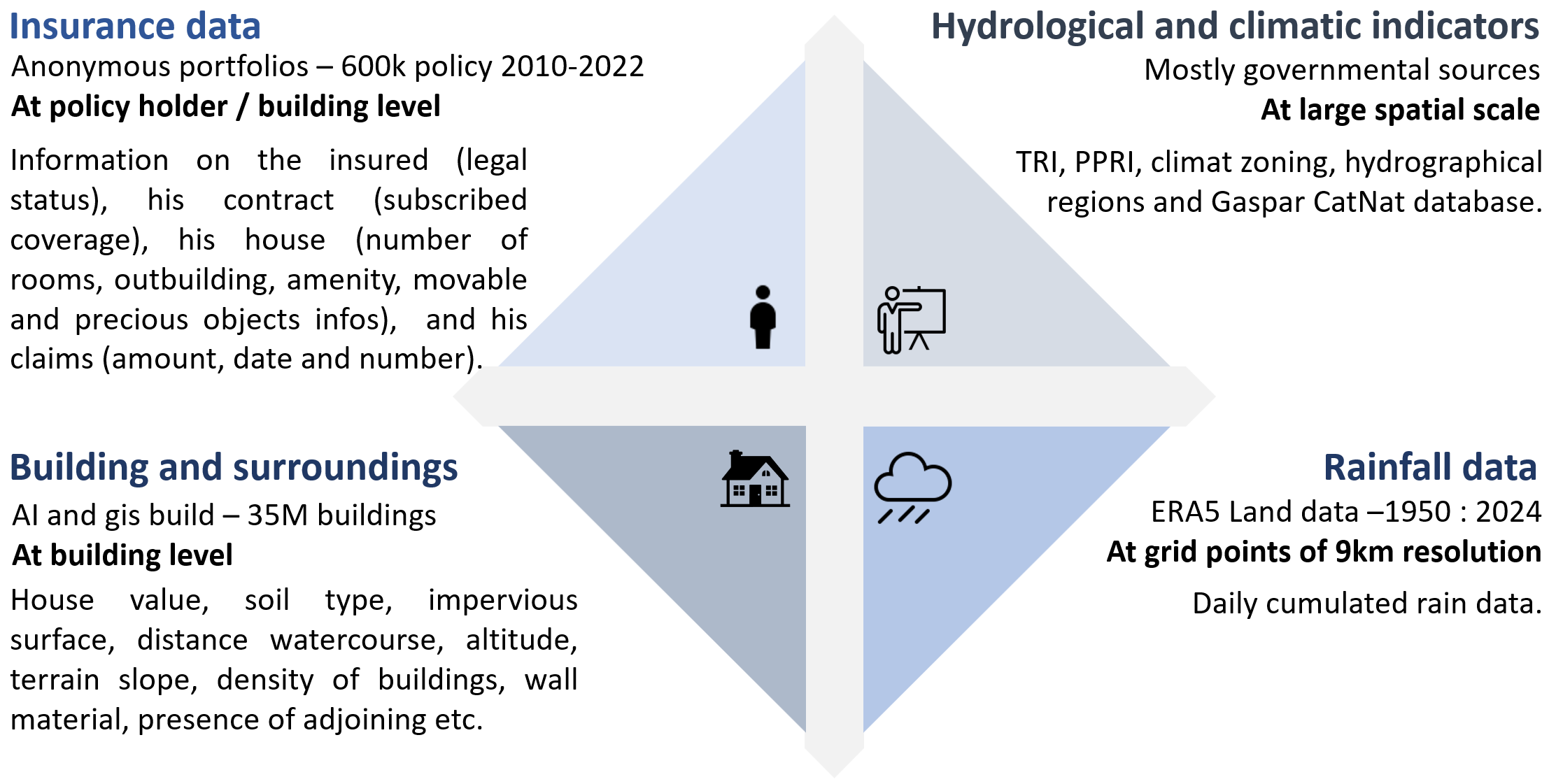}
    \caption{A summary of data sources used in this study. Building, hydrological and climatic data rely on the latest available version of the data source. The production years vary from a variable to another.}
    \label{fig:data_prez}
\end{figure}

This study combines several data sources to model both the occurrence and severity of flood claims at the building scale. The data comprise (i) an insurer’s portfolio and claims records, (ii) publicly accessible hydrological and climatic datasets, and (iii) georeferenced information characterizing the insured buildings and their surrounding context. Figure \ref{fig:data_prez} summarizes these data layers, detailing their origins and spatial resolutions.
Most of the external data sources used in this study are publicly accessible. Starting from these raw open datasets, we build and engineer a range of meteorological, environmental, and building-level variables that are tailored to fit standard insurer modelling frameworks. From a data point of view, our main contribution is to determine which variables can be derived from these open sources, to describe how they can be built, and to assess their respective added value for building-level flood-risk assessment. Although national disaster databases such as GASPAR \cite{gaspar2023} offer a broad coverage of flood events, they only document ministerial recognitions of natural disasters within the French Catastrophes Naturelles regime. Since recognition depends on a formal request from municipalities, smaller, highly localised, or unreported flood events are often missing. Furthermore, GASPAR’s spatial resolution is restricted to the municipality scale, which is too coarse for modelling at the building level. For these reasons, we rely on insurer portfolio data to evaluate the complete modelling framework and resort to aggregated national datasets only when examining the large-scale behaviour and internal consistency of rainfall-related variables.

Most of the result maps concentrate on three French regions, Occitanie, Hauts-de-France, and part of Grand Est, as illustrated in Figure \ref{fig:score_tail_cluster1d}. These regions exhibit contrasted flood-risk profiles. Occitanie is mainly influenced by a Mediterranean climate, with short-lived but highly intense rainfall and complex terrain, which together promote rapid runoff and flash flooding, especially during Cévenol-type events. By contrast, Hauts-de-France is characterised by a temperate oceanic climate, relatively flat landscapes, and flood mechanisms more frequently governed by slow river overflows and drainage-system saturation in low-lying zones. The selected departments in Grand Est correspond to intermediate configurations with respect to climate, hydrography, and topography. Apart from visualisation purposes, all analyses and performance metrics reported in this paper are computed from the complete national dataset.

The remainder of this section describes each data source and the derivation of the explanatory variables. Section \ref{sec:insurance_data} presents the insurance portfolio and claims dataset. Section \ref{sec:24expoclimat} discusses the hydrological and climatic indicators typically used in insurance applications. Section \ref{sec:rainfall_vars} introduces the variables derived from rainfall, including the construction of the MILRE indicator. Section \ref{sec:building} details the geolocated building-related and surrounding variables. Additional details on data sources and variable construction are provided in \ref{appendixA}. 

\subsection{Insurance data}\label{sec:insurance_data}
This study relies on data extracted from a large French home insurance portfolio covering the period 2014-2022. During the study period, the portfolio comprises approximately $968{,}000$ unique policyholders, with many policies spanning multiple years, and includes $10{,}800$ flood-related claims. The dataset includes residential insurance policies exposed to flood risk and associated claims. Each policy is geolocated at the building address level, enabling the integration of external geospatial information. 

For each insured building, the available underwriting information includes standard policy characteristics commonly used in insurance pricing, such as occupancy type, insured value, coverage options, and contract duration. These variables constitute the core information set traditionally available to insurers when modelling flood risk using statistical frequency–severity approaches. Claims data include the occurrence of flood related claims and the corresponding indemnity amounts. Claim costs are expressed in euros and adjusted to constant prices to account for inflation during the study period. 

The modelling framework distinguishes between two complementary targets: flood claim occurrence, defined as the probability that a policy experiences at least one flood related claim over a given exposure period; flood claim severity, defined as the total cost of a flood claim conditional on its occurrence. This separation is consistent with standard actuarial practice and allows different drivers to influence the probability and magnitude of losses. 

Insurance contract heterogeneity is a well-known issue when exploiting claims data (see \cite{andre_2013}). To ensure dataset homogeneity, we restrict the analysis to a sub-population of policies with comparable coverage and guarantees. This selection is based on contract-related information and is performed in collaboration with insurer data experts. Contract variables used for this filtering step are not included in the modelling phase, allowing the analysis to focus on the drivers of flood-risk rather than the contractual effects.

\subsection{Hydrological and climatic indicators}\label{sec:24expoclimat}

To characterise flood hazard using information typically available to insurers, we draw on a set of hydrological and climatic indicators constructed from publicly accessible datasets. These indicators are consistent with the reference flood risk evaluations produced by French public authorities.

Flood hazard zoning is primarily captured by the Territoires à Risque Important d’Inondation (TRI) and the Plans de Prévention du Risque Inondation (PPRI). The TRI, created under the European Directive 2007/60/EC (\cite{CEPRI2022floodlaw}), delineates areas exposed to significant flood risk, based on detailed hydraulic and topographic studies carried out for selected major rivers and coastal stretches. These zones are mapped at high spatial resolution and are intended to identify locations where flooding could generate major impacts on both population and economic activities. However, due to their targeted scope and the high cost of their development, TRI analyses cover only 5\% of the buildings in our Portfolio.
In contrast, PPRI (\cite{PPRI2023}) are regulatory planning tools defined at the municipal scale. They divide municipal areas into broad hazard zones using information from historical flood events, simplified hydraulic simulations, and expert assessment. Although PPRI covers a large part of the territory (approximately 30\% of insured buildings), their spatial granularity is relatively coarse and the delimitation of zones is often driven by administrative and land-use criteria rather than detailed representations of flood processes.

Beyond TRI and PPRI, we also incorporate additional geographical descriptors related to hydrography and climate, such as river basin boundaries and climatic regions. These features represent the long-term environmental setting and capture broad contrasts in precipitation regimes across France, but they are either static or evolve only slowly over time.

Together, these hydrological and climatic variables constitute the standard open source baseline hazard information commonly used by insurers. They offer an institutional reference for flood risk evaluation, yet they are fundamentally constrained in their capacity to represent the fine-scale spatial and temporal variability of rainfall-induced flooding at the individual-building scale. In this work, they are used both as explanatory covariates and as reference to assess the incremental value of rainfall-based metrics and building-level attributes.

\subsection{Rainfall related variables}\label{sec:rainfall_vars}

\begin{figure}[h]
     \centering
     \includegraphics[width=0.8\textwidth]{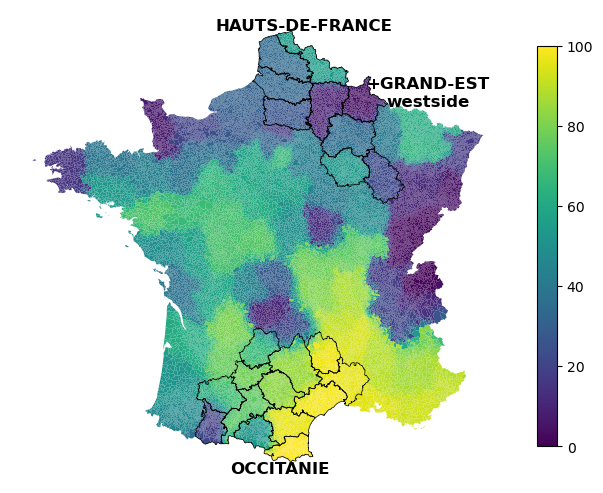}
     \caption{Estimated tail weights with the shape parameter of a generalised Pareto distribution, rescaled between $0$ and $100$, where $100$ corresponds to the heaviest tails. Selected interest zones for modelling results visualization are also delineated in black.}
     \label{fig:score_tail_cluster1d}
\end{figure}

Flood losses are often triggered by intense or persistent rainfall events, but the relationship between precipitation and resulting damages is indirect and highly nonlinear. Flooding may occur through multiple mechanisms, including surface runoff, river overflow, soil saturation, or drainage-system failure, and damages are often caused by the horizontal redistribution of rainfall rather than by precipitation intensity at a single point in time. As a result, raw rainfall measurements are rarely suitable as direct predictors of flood occurrence or cost at the building level. 

To better capture these dynamics within an insurance modelling framework, we construct rainfall-related variables that account for both spatial heterogeneity in extreme rainfall behaviour and the temporal structure of rainfall events. All rainfall variables are derived from the ERA5-Land reanalysis dataset, which provides daily precipitation estimates on a regular grid covering metropolitan France, with approximately 6200 grid points and observations spanning from 1950 to 2024 (\cite{ECMWF}).

\vspace{3mm}

\textbf{Spatial characterisation of rainfall extremes}: extreme rainfall distributions exhibit strong spatial variability across France, reflecting contrasted climatic regimes. To account for this heterogeneity, we estimate the tail behaviour of daily rainfall distributions using extreme value theory. Specifically, we fit a generalised Pareto distribution to exceedances over high thresholds and use the estimated shape parameter as a proxy for tail heaviness (\cite{coles2001introduction}). Grid points with similar tail behaviour are then grouped using a clustering approach, leading to the construction of a categorical variable denoted \texttt{tail\_weight\_cluster}. Figure \ref{fig:score_tail_cluster1d} presents the raw tail weights before clustering, rescaled between $0$ and $100$, where $100$ corresponds to the heaviest tails. The resulting patterns are consistent with known climatic contrasts, with heavier-tailed rainfall distributions in southern regions and milder tails in northern areas.

\vspace{3mm}

\textbf{Event-based rainfall aggregation :} in addition to spatial heterogeneity, flood occurrence is strongly dependent on the temporal structure of rainfall. Some events are driven by short and intense precipitation, while others result from cumulative rainfall over several days. To capture these mechanisms, we compute rainfall aggregates over multiple time windows and extract, for each exposure period, the most extreme accumulation across windows.
These constructions are motivated by hydrological considerations and aim to approximate key drivers of flood generation, such as soil saturation and runoff potential, while remaining independent of explicit hydraulic modelling.

\vspace{1mm}

We introduce a synthetic rainfall indicator, referred to as the Most Intense Local Relative Event (MILRE), designed to jointly address spatial heterogeneity and multi-scale temporal accumulation. For flood-severity modelling, each claim is associated with a known flood date. From this date backward, the depths of the accumulated rainfall are computed for several time windows preceding the event, namely $nd \in \{1, 3, 5, 7, 10, 30\}$ days. For each insured building, rainfall values are obtained from the four closest points on the ERA5-Land grid, and we retain the maximum value over these points to represent local precipitation conditions. This value is noted $i_{nd}$. 

\begin{figure}[!h]
     \centering
    \includegraphics[width=\textwidth]{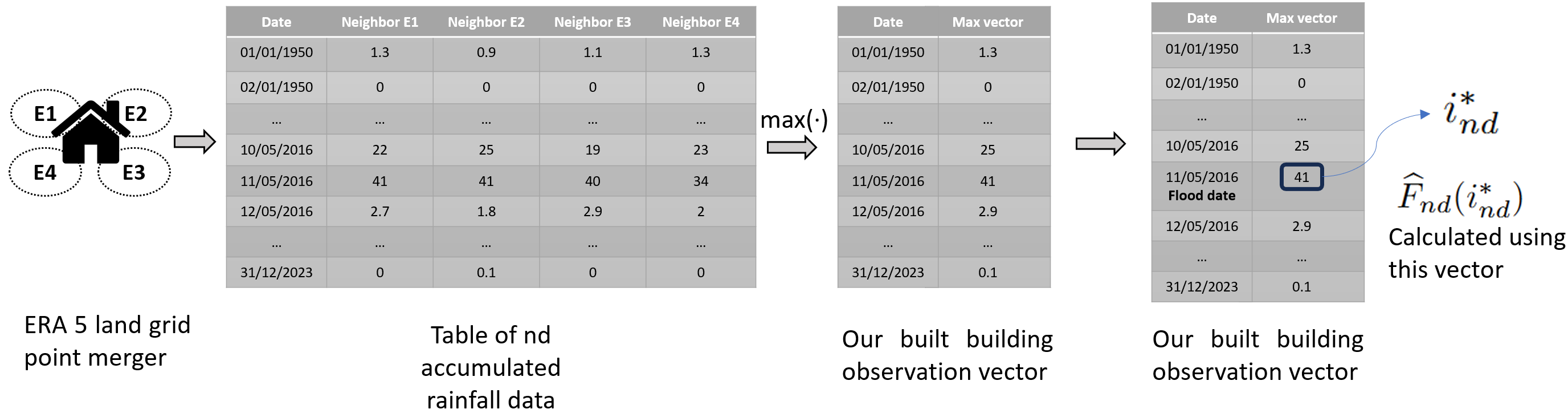}
    \caption{Schematic description of the process needed to obtain the empirical probability associated with a building, for a given time window and flood date.}
    \label{fig:illus_q_precip}
\end{figure}

For each accumulation window $nd$, the observed rainfall value on the flood date is transformed into an empirical cumulative probability, estimated from the historical distribution of the corresponding accumulated rainfall series at that location :  $$\widehat F_{nd}(i_{nd}^*) = \frac{1}{M_{nd}}\sum_{k=1}^{M_{nd}} \mathbbm{1}_{i_{nd}^k\leqslant i_{nd}^*},$$ where $M_{nd}$ is the number of days in our vector and $i_{nd}^*$ the observed value associated with the flood event. We move from absolute rainfall depth value in millimetres to empirical probabilities as a practical way to express value in relative terms, facilitating comparison between time windows and also across areas with different rainfall behaviours.
Figure \ref{fig:illus_q_precip} is a schematic description of the process needed to get $\widehat F_{nd}(i_{nd}^*)$ associated with a building, for a given time window and flood date. 

MILRE is then defined as the maximum empirical probability observed across all accumulation windows,

\begin{equation}\label{eq:MILRE_1_bis}
   MILRE =  \widehat F_{max} = \max_{nd \in \{1, 3, 5, 7, 10, 30\}}\widehat F_{nd}(i_{nd}^*).
\end{equation}

By construction, MILRE (Most Intense Local Relative Event) takes values between 0 and 1, with higher values indicating rainfall events that are extreme \textbf{relative} to local historical conditions. The use of multiple accumulation windows ensures that we get the  \textbf{most intense event} from both short-duration rainfall and long-term cumulative effects.
\vspace{3mm}

\begin{figure}
 \begin{subfigure}{0.49\textwidth}
     \includegraphics[width=\textwidth]{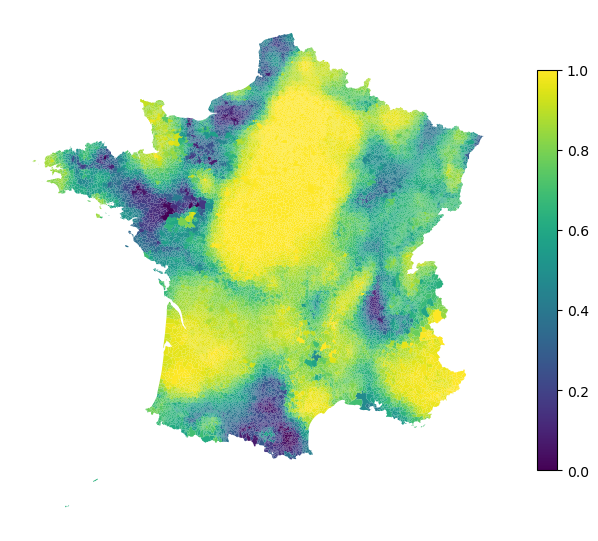}
     \caption{\texttt{ann\_MILRE} values for year 2016.}
     \label{fig:MILRE_2016}
 \end{subfigure}
 \hfill
 \begin{subfigure}{0.49\textwidth}
     \includegraphics[width=\textwidth]{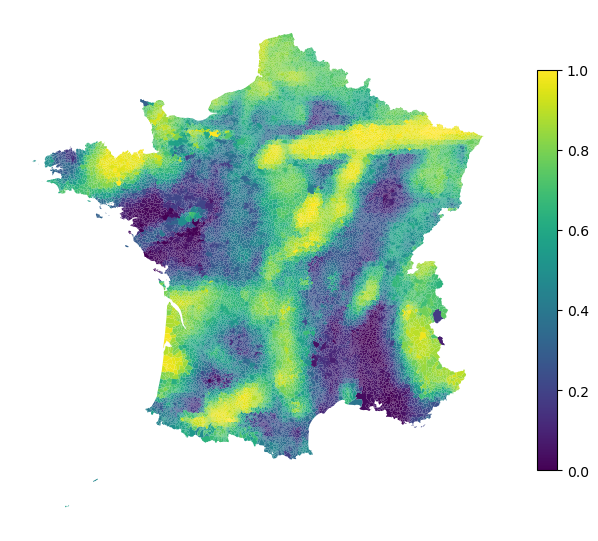}
     \caption{\texttt{ann\_MILRE} values for year 2017.}
     \label{fig:MILRE_2017}
 \end{subfigure}

 \caption{\texttt{ann\_MILRE} (annual Most Intense Local Relative Event) spatial distribution for 2016 and 2017. Higher values indicate regions that have undergone extreme events relative to their historical rainfall levels.}
 \label{fig:annual_MILRE}

\end{figure}

\textbf{Annual MILRE for occurrence modelling} : for the frequency dataset, using flood dates is irrelevant. Properties are covered for some part of the year, and we know how many flood related claims were submitted. 

An analogous annual indicator, denoted \texttt{ann\_MILRE}, is thus defined by selecting, for each year and each accumulation window, the maximum rainfall recorded during that year, converting this value into an empirical probability and then maintaining the maximum over all windows.
More precisely, for a given cumulative time window $nd$ (recall: 1 day, 3 days, 5 days, etc.), we first determine the annual maximum of the corresponding accumulated rainfall from 1950 to 2024. This yields $nd$ vectors, each containing the annual maxima. We then compute, for each of these vectors, the associated empirical cumulative distribution, following the same procedure used for $i_{nd}^*$ and $F_{nd}(i_{nd}^*)$. Conceptually, this provides, for each year, $nd$ values that represent the empirical cumulative probability of annual maxima. We thus obtain : 

$$\widehat F_{nd}(y_{nd}^*) = \frac{1}{75}\sum_{k=1950}^{2024} \mathbbm{1}_{y_{nd}^k\leqslant y_{nd}^*},$$  
where $y_{nd}^*$ denotes the annual maximum observed for the accumulation window $nd$ in the specific year under consideration.

Finally, we combine these vectors into a single vector taking, for each year, the maximum of the $nd$ probability values as in \ref{eq:MILRE_1_bis} : 
\begin{equation}\label{eq:ann_MILRE}
   ann\_MILRE = \max_{nd \in \{1, 3, 5, 7, 10, 30\}}\widehat F_{nd}(y_{nd}^*).
\end{equation}

This variable indicates whether a particular year saw extreme rainfall events and, if so, identifies their locations. The basic idea is that it helps exclude regions and years that did not experience substantial rainfall-related extremes.
In Figure \ref{fig:annual_MILRE}, which shows the ann\_MILRE maps for 2016 and 2017, the contrast in the distribution of high values is particularly evident. The 2017 map displays far fewer high-intensity events than 2016. The year 2016 is widely recognized for its severe flooding episodes in France (\cite{juin20162023}), whereas 2017 is characterized by a larger number of areas with very low values, indicating a lack of extreme rainfall events. While some regions in 2016 also did not experience substantial rainfall, these zones are less spatially widespread than in 2017. To quantify the discrepancy in flood occurrences between 2016 and 2017, we refer to the Gaspar CatNat event counts. In the GASPAR extract used in this study (downloaded in 2022), 2016 ranks as the 8th highest year in terms of the number of municipalities officially declared as having been affected by flooding, whereas 2017 ranks only 41st.

\begin{figure}[!h]
 \begin{subfigure}{0.49\textwidth}
     \includegraphics[width=\textwidth]{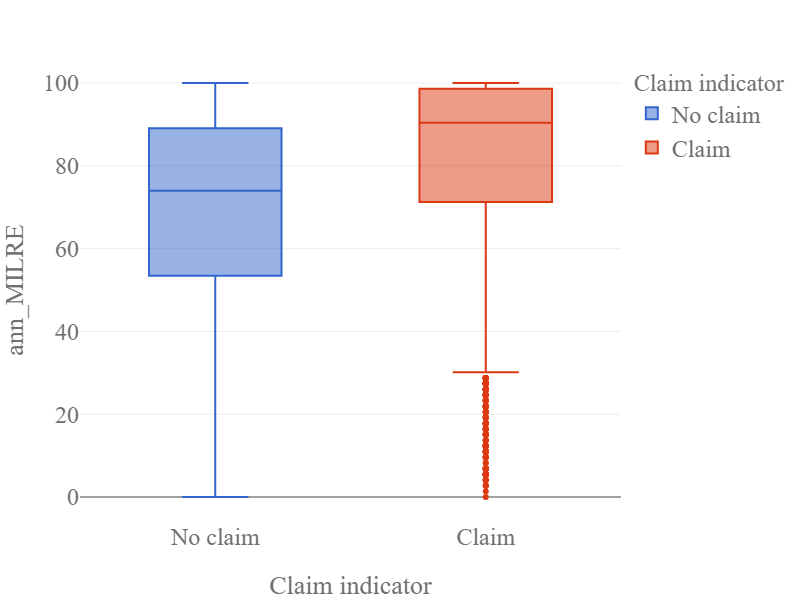}
     \caption{\texttt{ann\_MILRE} boxplot for the insurance portfolio dataset filtered by claim occurrence.}
     \label{fig:boxCLM}
 \end{subfigure}
 \hfill
 \begin{subfigure}{0.49\textwidth}
     \includegraphics[width=\textwidth]{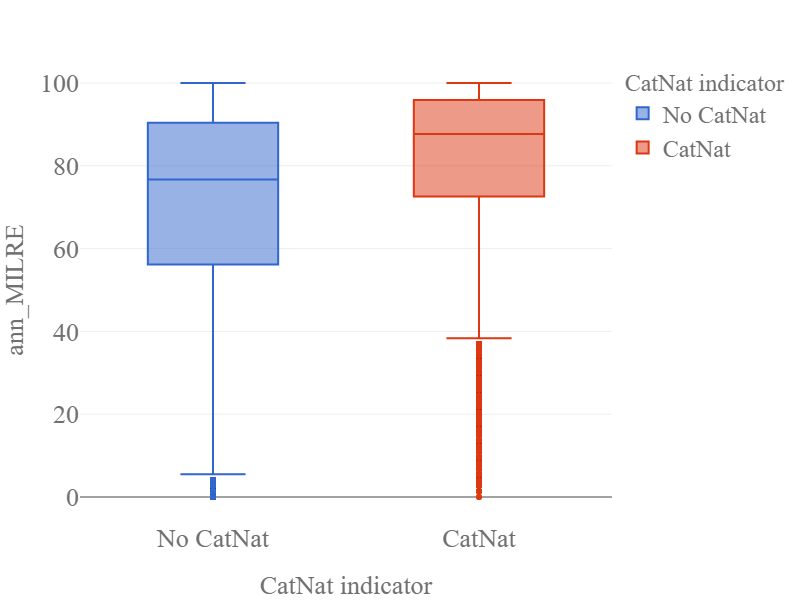}
     \caption{\texttt{ann\_MILRE} boxplot for GASPAR dataset filtered by presence of at least one accepted CatNat decree.}
     \label{fig:boxGAS}
 \end{subfigure}

 \caption{Box plots of \texttt{ann\_MILRE} by flood occurrence for the GASPAR and claim datasets, illustrating distributional differences between periods with and without events.}
 \label{fig:box_annual_MILRE}

\end{figure}

In Figure \ref{fig:box_annual_MILRE}, we analyse how the distribution of \texttt{ann\_MILRE} values varies depending on whether a claim was filed and whether a CatNat decree was issued. Subfigure \ref{fig:boxCLM} compares \texttt{ann\_MILRE} values between contracts (buildings) that reported a claim in a given year and those that did not. Subfigure \ref{fig:boxGAS} displays boxplots conditioned on the presence of a CatNat decree, using open data from the GASPAR dataset. This dataset provides, for each year, the list of municipalities for which a CatNat decree was issued, together with the associated event dates. For each year, we extend this list by adding municipalities that do not appear in the dataset, so that all French municipalities are represented each year. We then plot boxplots of the corresponding \texttt{ann\_MILRE} values for each category.

Across both subfigures, the orange box indicates values clustered around $100\%$, showing that when a claim or decree occurs, our variable is generally higher than when there is no flood event. Most values exceed $35\%$, which is not the case in the absence of an event. Although the distributions partially overlap, they remain clearly distinct. Reaching similar conclusions on the GASPAR dataset is a noteworthy result. Its temporal and spatial coverage is broader than that of the claims in our portfolio (1983–2023 for all French municipalities).

\subsection{Building and its surroundings} \label{sec:building}

Flood losses are heavily shaped by local vulnerability conditions that are not reflected by hazard maps or rainfall metrics alone. These conditions concern both the inherent properties of buildings and the characteristics of their immediate surroundings, which together determine how floodwaters reach insured structures and how damage develops. In contrast to perils such as hail, where the damaging agent strikes the building directly, flooding is an indirect hazard: rainfall initiates a chain of processes involving terrain, land use, and the built environment before ultimately producing damage.

To represent these influences, we assemble a set of geolocated variables that describe insured buildings and their local contexts. These variables are derived from open-source geographic data and are spatially linked to insured locations at the individual-building scale. They are intended to augment insurers’ underwriting data while remaining consistent with typical pricing practices and established flood risk assessment procedures.

\textbf{Building characteristics} : building-level variables describe structural attributes that influence vulnerability to flooding and the magnitude of potential losses. These include footprint area, construction type, number of floors, and primary wall materials. Some of these characteristics are already used in insurance pricing and underwriting. When available within insurer systems, they also contribute to improved data consistency by supporting geocoding validation, detecting reporting errors or omissions, and improving the overall reliability of exposure information. In this study, building characteristics are used to capture relative differences in vulnerability across insured assets rather than to represent detailed structural or engineering properties. Their role is to provide additional contextual information at the building scale that complements insurer underwriting data and hazard indicators.

\textbf{Surrounding context (built and natural)} : the local environment plays a critical role in flood generation, propagation, and concentration at the local scale. Therefore, we include variables describing land use, soil sealing, topography, and hydrographic context in the vicinity of insured buildings. These indicators encompass measures such as surface imperviousness, vegetation coverage, local slope, relative elevation, and distance to watercourses. Together, these variables determine the intensity of runoff, the extent of flow accumulation, and whether rainfall can infiltrate or is instead rapidly drained away. They are calculated within predefined spatial neighbourhoods surrounding each building, providing a concise and practical summary of local conditions without explicitly simulating flow paths. Although factors such as evapotranspiration and flood protection infrastructure can also affect flood dynamics, consistent and spatially comprehensive data on these aspects are not available at the national level. Consequently, they are not explicitly incorporated into the set of environmental variables used in this study.

The contribution of building-level and environmental variables is evaluated incrementally in the modelling framework, both individually and through interactions with rainfall-related variables, to assess how the local context modulates the impact of extreme precipitation on flood occurrence and severity.

\vspace{3mm}
We would like to conclude this section by stating that France Assureurs projects that flood-related losses will increase by $81\%$ by 2050 \cite{assureurs2021impact}. Of this increase, $34\%$ is linked to climate influences, including $13.77\%$ directly attributable to climate change. For example, \cite{luu2018attribution} reports that in southern France, the probability of exceeding the centennial flood amplitudes has increased by $2.5(\pm 0.8)$ percent. Further evidence from attribution studies can be found in \cite{philip2018validation,bloschl2019changing}. Together, these results highlight the need to integrate meteorological and hydrographic indicators into flood risk models, both to represent the main drivers of risk and to track how flood hazards change under a warming climate. In particular, France Assureurs uses the $90\%$ and $99\%$ quantiles of daily rainfall to explain and forecast flood losses, suggesting that a richer use of such variables in insurance analyses would improve our understanding and prediction of flood damage.

France Assureurs also attribute $65\%$ of the anticipated cost increase to a ``wealth effect'', reflecting growing urbanisation and higher value of assets. This emphasises the relevance of examining household behaviours and settlement patterns in climate risk assessments (\cite{bezy2023incidence}). Although this dimension lies beyond the scope of our work, it further motivates linking claims with detailed building attributes and local environmental conditions (\cite{torgersen2017evaluating}).

\noindent Table \ref{tab:data_overview} gives an overview of the variables considered for modelling in this study.  



\begin{table}[!ht]
\centering
\resizebox{\textwidth}{!}{%

\begin{tabular}{l p{6.5cm} p{2.5cm} p{3.5cm}}
\cmidrule{2-4}

 ~~ & \textbf{Variables} & \textbf{Spatial resolution} & \textbf{Comment} \\
\cmidrule{2-4} 

\textbf{\underline{Insurance data}}&  &   &    \\ [0.2cm]

 ~~& claim occurrence, claim amount & building & modelling targets \\ [0.3cm]
 & number of rooms, movable assets, precious objects, outbuilding size & building & underwriting information \\
 
 \textbf{\underline{\makecell[l]{Hydrological and \\climatic data}}}&  &   &    \\ [0.4cm]

 ~~& \makecell[l]{PPRI, climatic regions,\\ number of CATNAT decrees} & municipality & \makecell[l]{regulatory and historical\\ expert zoning} \\ [0.3cm]
 
 & TRI, catchment zones & refined polygons & \makecell[l]{regulatory and historical\\ expert zoning} \\
 
 \textbf{\underline{Rainfall data}}&  &   &    \\ [0.2cm]

 ~~& extreme tail weight clusters & municipality &  \makecell[l]{spatial characterisation of\\ rainfall extremes}\\ [0.3cm]
 
 & MILRE, ann\_MILRE & \makecell[l]{ERA5 grid recalculated \\ at the building level} & \makecell[l]{event-based rainfall \\intensity indicators}   \\

 \textbf{\underline{\makecell[l]{Building and \\surrounding data}}}&  &   &    \\ [0.4cm]

 ~~& \makecell[l]{living surface, house value, construction period,\\ number of floors, wall material, exterior building surface,\\ total wall length, presence of adjoining building} & building & \makecell[l]{structural and vulnerability\\ descriptors} \\[0.6cm]
 
 ~~& \makecell[l]{terrain max slope, building density, \\predominant soil type, impervious surface percentage} & \makecell[l]{buffers around \\ the building} & local environmental context \\[0.4cm]

 ~~& distance and altitude difference to the closest watercourse & building & proximity to hydrographic network  \\

\cmidrule{2-4} 
\end{tabular}}

\caption{Overview of variable groups, spatial resolution, and modelling role. Detailed variable definitions and descriptive statistics are provided in Appendix~A.}
\label{tab:data_overview}
    
\end{table}

\newpage

\section{Modelling flood claim cost and probability framework}\label{sec:modeling}

\subsection{Modelling approach}\label{sec:modelingapp}
For each insured $i \in {1,\dots,n}$, let $\mathbf{Y_i}$ denote the target variable, which typically represents the frequency of the claim, probability of occurrence, severity, or pure premium in casualty insurance. Let $N_i$ be the number of claims and $C_{i,j}$ the cost of claim $j$. In practice, insurers model the occurrence and severity separately using different datasets (see \cite{goldburd2016generalized}). In cases where at most one claim per year is observed per insured, as in our dataset, the total expected cost (or pure premium) can be modelled as: $$\mathbb{E}(Y_i) = \underbrace{\mathbb{E}(\mathbbm{1}_{N_i>0})}_\text{probability module}  \times ~~~~\underbrace{\mathbb{E}(C_{i,1})}_\text{severity module}.$$

The traditional pricing approach is based on generalised linear models (GLM) since \cite{nelder1972generalized,brockman1992statistical}. This choice is deliberate: our objective is not to propose new modelling techniques, but to evaluate how additional contextual and geolocated variables improve flood-risk modelling within an operational insurer setting. In our setting, we use logistic regression (a GLM with Bernoulli distribution and logit link) to model claim probability, and a log-linked Gamma GLM for claim cost. In both cases, the linear predictor takes the form of : 

\begin{equation}\label{eq:glm_}
g(\mathbb{E}(Y_i)) = \beta_0 + \beta_1 x_{i,1} + \beta_2 x_{i,2} + \cdots + \beta_p x_{i,p} ,\end{equation} where $g(\cdot)$ denotes the link function, $(x_{i,1},\dots,x_{i,p})$ are explanatory variables, and $(\beta_0,\dots,\beta_p)$ are the associated coefficients.

The first modelling layer relies exclusively on standard underwriting information and serves as a baseline. Subsequent layers progressively incorporate additional data sources, including expert climatic and hydrological flood zoning variables, rainfall indicators, and building and surrounding characteristics. All models presented in the main text follow the same GLM structure and are estimated using identical training and evaluation protocols. Improvements in predictive performance are therefore directly attributable to the inclusion of new data layers rather than to changes in modelling techniques.

Spatial heterogeneity plays a central role in flood risk, and insurers often rely on zoning variables derived from administrative or expert sources to account for geographical differences in exposure. Several studies propose two-step approaches in which non-spatial models are first estimated and residual spatial structure is then captured through latent zoning variables or spatial smoothing techniques before being reintegrated in the pricing model (see \cite{taylor1989use,boskov1994premium,Assuncao2014,scheel2013bayesian}). In this study, spatial information is introduced primarily through explicit covariates. This choice reflects our objective of directly linking the occurrence and severity of floods to interpretable physical and contextual drivers at the building level. 

\subsection{Fitting details and model evaluation}

To assess model performance and validity, multiple evaluation metrics are employed, each customised to the specific modelling challenge. The cost model is evaluated using the following metrics: 

\begin{itemize}
    \item Root mean square error (RMSE) $\Big[\frac{1}{n}\sum_{i=1}^n(\widehat y_i-y_i)^{2}\Big]^{1/2}$ penalises large prediction errors and provides an absolute measure of predictive accuracy. Lower values indicate better performance.
    \item Gini index evaluates the model’s ability to rank risks from the least to the most important (see \cite{frees2014insurance,denuit2019model}).
    
     The predictions are ordered as $\widehat y_{(n)}\geq \dots \geq \widehat y_{(1)}$, and the cumulative proportions of exposure $P_{(i)} = i/n$ and the total observed cost $R_{(i)} = \sum_{j=1}^{i} y_j / \sum_{j=1}^{n} y_j$ are calculated based on this ordering. The index is derived from the associated Lorenz curve using the standard formula \cite{dorfman1979formula}. Higher values indicate a stronger ranking power.
    
    \item Deviance ~~$2\sum_{i=1}^n \Big[(y_i - \hat{y}_i)/\hat{y}_i - \log(y_i/\hat{y}_i)\Big]$ which compares the log-likelihood of a fitted model to the log-likelihood of a saturated model, which perfectly predicts the observed data. The lower the value, the better the model.  
\end{itemize}

Claim occurrence modelling is characterised by extreme class imbalance, with approximately $0.22\%$ positive outcomes. In such settings, standard training and evaluation procedures can lead to biased models that underestimate rare events \cite{he2009learning}. To address this issue, we adopt a class re-weighting strategy during GLM estimation \cite{wang2017learning,zhu2018class}. Observation weights are defined as:

$$w_0 = \frac{n}{2*\sum_{i=1}^n \mathbbm{1}_{y_i=0}} \text{ and } w_1 = \frac{n}{2*\sum_{i=1}^n \mathbbm{1}_{y_i=1}},$$
$w_0$ and $w_1$ are, respectively, weights for the majority and minority classes. This choice preserves the total effective sample size while penalising misclassification of rare events more heavily. It can be viewed as a simplified variant of cost-sensitive learning, because only one type of error penalty is altered.

In imbalanced classification problems, traditional metrics, such as accuracy, are often misleading (see \cite{guo2004learning,japkowicz2013assessment}). Instead, we rely on: 

\begin{itemize}
    \item Log loss (-1/2n deviance): $-\frac{1}{n}\sum_{i=1}^N\Big[(1-y_i)\log(1-\widehat y_i)+y_i\log(\widehat y_i)\Big]$, which evaluates probabilistic calibration. Logloss is proportional to deviance in the case of logistic regression. We report -logloss in the results table; smaller values indicate better performance.
    \item Gini index used to assess ranking ability, which is more informative than ROC-based metrics under severe imbalance \cite{jeni2013facing}.
    \item Critical success index: $\textit{true positive}/(\textit{true positive}+\textit{false positive}+\textit{false negative})$ introduced in \cite{schaefer1990critical} to measure the skill of a classifier in an extreme imbalanced setup. See \cite{legrand2021evaluation} for more recent use and discussions on this metric. Contrary to probabilistic or ranking metrics, this metric is defined with predicted class labels and not probabilities. To enable a fair comparison across models, we determine for each model the threshold that produces the maximum CSI value and report this maximum as the metric. A higher metric value indicates better performance. Using this threshold and the corresponding predicted class,  we also calculate the classification precision and recall for the flooded individuals. 
\end{itemize}

All models are evaluated using 5-fold cross-validation. In addition to absolute performance metrics, results are reported relative to a baseline insurer model using underwriting data only. A dummy model without covariates is also included as a minimal benchmark. Model parsimony is assessed by reporting the number of estimated parameters.

To assess the contribution of individual covariates, we perform likelihood ratio tests by comparing the full model with restricted models excluding one variable at a time. Under the null hypothesis that the excluded variable does not improve the model, the test statistic follows a chi-square distribution. The results are reported on a logarithmic scale using a $5\%$ significance threshold. For selected variables of interest, observed-versus-predicted plots are provided to illustrate their marginal impact on risk estimation. We also report the number of model parameter to evaluate parsimony.

\noindent The following modelling results' sections are structured in layers: 
\begin{enumerate}
    \item GLM ins: baseline modelling with underwriting variable only.
    \item GLM ins+c: extension of the baseline model with expert-based hydrological and climatic variables.
    \item GLM ins+r: further enhancement including rainfall-derived indicators in addition to hydrological and climatic variables.
    \item GLM all: full model incorporating rainfall, climate expertise, building, and surrounding variables.
\end{enumerate}

\subsection{Cost modelling results and discussions}\label{sec:cost_res}
\begin{figure}[!h]
     \centering
    \includegraphics[width=\textwidth]{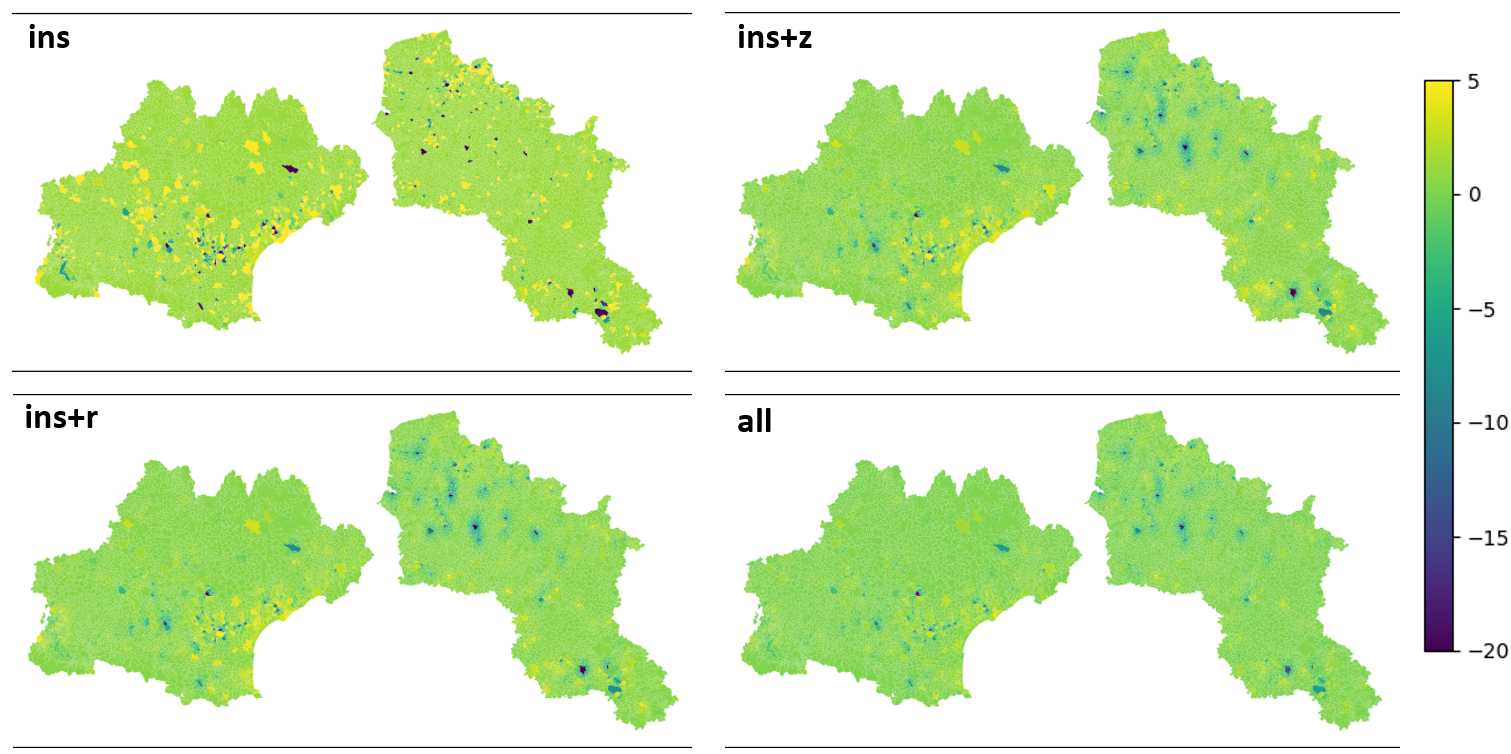}
    \caption{For each of four GLM cost models defined in Table \ref{tab:res_cost_mod} and the two regions defined in Figure \ref{fig:score_tail_cluster1d}, Pearson residuals are displayed at the municipality level. A strong departure from zero indicates a misfit between observed and fitted values.}
    \label{fig:dep_costres_agg}
\end{figure}

\newcommand{\formatvarv}[2]{\makecell[c]{#1\\\emph{\tiny{#2}}}}
Table \ref{tab:res_cost_mod} reports the evaluation metrics for each model. The model names shown in this table are used consistently across maps and plots to support interpretation. Figures \ref{fig:dep_costres_agg} display the Pearson residuals for all models. Residual maps are generated by aggregating values at the municipality scale using the median within each group, whereas performance metrics are calculated at the individual building level. While this aggregation unavoidably reduces some detail, it offers the best compromise between clear visualization of the results and preserving sufficient granularity to examine the spatial patterns of our models.

\begin{table}
\centering
\resizebox{0.9\textwidth}{!}{%
\begin{tabular}{m{3.0cm}m{1.5cm}m{1.5cm}m{1.5cm}m{1.5cm}}
\toprule
\makecell{Models} & \makecell{RMSE} & \makecell{Gini} & \makecell{Deviance} & \makecell{Nb params}\\ \midrule
 \makecell{\textbf{Dummy model}\\ \emph{\small{no variable}}} & \formatvarv{9165.4983}{\textcolor{red}{$\shneg0.21\%$}} & \formatvarv{0}{\textcolor{red}{$\shpos100\%$}} & \formatvarv{13340.42}{\textcolor{red}{$\shneg1,21\%$}} & \formatvarv{\textbf{0}}{\textcolor{blue}{$\shpos100\%$}}\\ [0.3cm] 

 \makecell{\textbf{GLM ins}\\ \emph{\small{insurer data only}}} &  \makecell{9146.6254} & \makecell{0.0562} & \makecell{13180.49} & \makecell{15}\\ [0.3cm] 
 
 \makecell{\textbf{GLM ins+c} \\ \emph{\small{insurer data + climate}}} & \formatvarv{9054.3267}{\textcolor{blue}{$\shpos1.01\%$}} & \formatvarv{0.1474}{\textcolor{blue}{$\shneg162.28\%$}} & \formatvarv{12586.02}{\textcolor{blue}{$\shpos4.51\%$}} & \formatvarv{19}{\textcolor{red}{$\shneg26.67\%$}} \\[0.4cm]
 
  \makecell{\textbf{GLM ins+r} \\ \emph{\small{insurer data + rainfall}}} & \formatvarv{9000.7446}{\textcolor{blue}{$\shpos1.59\%$}} & \formatvarv{0.1661}{\textcolor{blue}{$\shneg195.55\%$}} & \formatvarv{12424.03}{\textcolor{blue}{$\shpos5.74\%$}} &\formatvarv{19}{\textcolor{red}{$\shneg26.67\%$}} \\[0.4cm]
  
   \makecell{\textbf{GLM all} \\ \emph{\small{all data}}} & \formatvarv{\textbf{8781.6942}}{\textcolor{blue}{$\shpos3.99\%$}} & \formatvarv{\textbf{0.2439}}{\textcolor{blue}{$\shneg333.99\%$}} & \formatvarv{\textbf{11619.99}}{\textcolor{blue}{$\shpos11.84\%$}} & \formatvarv{33}{\textcolor{red}{$\shneg120\%$}}\\[0.4cm]\bottomrule
\end{tabular}}
\caption{Summary of cost models' metrics calculated with 5-fold cross validation and their relative evolution regarding the insurer data only model. Evolution is in blue when the metric is improving and in red if not.}
\label{tab:res_cost_mod}
\end{table}

\paragraph{\textbf{Insurer model, ``GLM ins'', in table \ref{tab:data_overview} insurance data only}}~~\\ \noindent The first key observation is that our insurer model, while slightly outperforming the dummy benchmark, is still relatively weak. It delivers only modest improvements in error measures and shows very limited risk differentiation, as indicated by a Gini index of $0.0562$ in table \ref{tab:res_cost_mod}. This result is unsurprising since insurance data mainly reflect the insured asset value, which corresponds to the traditional insurance product structure. 

In addition, the link between costs and house value is weak as flood-related damage is largely determined by hydrological drivers such as water volume, speed, and depth. The insurer model excludes intensity-related predictors and lacks indicators of whether a property is located in a high-risk flood area. Consequently, it does not establish strong relationships with claim costs. The variables that contribute the most to the model are the value of movable contents and those associated with the size and value of the building, such as the number of rooms. These help explain part of the variation in insured losses by scaling the estimated loss amount. However, the model does not yield any meaningful spatial segmentation, even though this is essential for climate-sensitive risks.

Previous research on insurance pricing has shown that underwriting data alone do not permit sufficiently precise risk segmentation. Insurers therefore frequently enhance their internal data with external sources, including geographic and environmental information (see \cite{denuit2004non}). In line with this, we extend the model by adding an additional layer that integrates climate expertise variables.

\paragraph{\textbf{Insurer + Climate model, ``GLM ins+c'', in table \ref{tab:data_overview} insurance + hydrological and climatic variables}} \noindent This modelling layer builds on the baseline specification by adding expert-defined climate and flood zoning variables, such as TRI, PPRI, climatic regions, hydrographic structures, and historical CatNat records. These are directly used as covariates and represent the kind of information insurers typically use to account for spatial differences in exposure when more sophisticated modelling approaches are not available. Among these expert-based risk variables, TRI categories and climatic region classes have the strongest impact on the insurance + climate cost model. TRI are delineated in areas that combine past exposure, specific hydrological conditions, and high potential economic losses, which explains their strong association with observed damages. Climatic regions provide complementary large-scale information by distinguishing areas based on rainfall regimes and dominant flood-generating mechanisms, thereby capturing persistent spatial contrasts in flood damage patterns and highlighting climates that are intrinsically more prone to flooding.

Adding the climate expertise variable leads to a $1\%$ reduction in RMSE and a threefold increase in the Gini index, pointing to better risk segmentation. As shown in Figure \ref{fig:dep_costres_agg}, spatial patterns begin to emerge, in particular along the Mediterranean coastline and the Garonne River in eastern Occitanie. In Hauts-de-France and Grand Est, the climate model sharpens risk detection in the southern parts of these regions, while segmentation in the northern areas remains limited. This gain is visible as a reduction in the residual errors on the map.

\paragraph{\textbf{Insurer + rainfall model, ``GLM ins+r'', in table \ref{tab:data_overview} insurance + Climate + rainfall variables}} ~~\\ \noindent
We now introduce indicators based on rainfall, focussing on extreme events through the MILRE variable, which captures the most intense local rainfall episode associated with each flooded building. MILRE is discretised into quantile-based classes to balance flexibility and robustness, but it enters the model as a continuous predictor. This variable improves the model’s ability to explain claim severity. The logic is straightforward: weaker rainfall events tend to generate smaller insurance payments, whereas extreme episodes are more likely to cause large losses. The empirical results in table \ref{tab:res_cost_mod} support this view: incorporating extreme rainfall information yields a $1.6\%$ reduction in RMSE and raises the Gini index to $16.6\%$, outperforming the climate-only model. This suggests that rainfall intensity contributes information that complements expert zoning by describing the severity of individual events rather than long-term exposure.

A comparison with the insurance + climate (\texttt{ins+c}) model shows that, while their large-scale residual structures are similar, the \texttt{ins+r} specification delivers more detailed and differentiated predictions (Figure \ref{fig:dep_costres_agg}). In northern departments and along the Occitanie coastline, for instance, it produces a smoother residual gradient instead of the uniformly high residuals observed in the \texttt{ins+c} model. This indicates that including extreme rainfall information refines the spatial resolution of risk estimates.

\paragraph{\textbf{All variables model, ``GLM all'', in table \ref{tab:data_overview} insurance + rainfall + climate + building variables}} \noindent
Finally, we build a fully integrated model that uses all available data sources, namely insurance records, meteorological variables, climate-related indicators, and the characteristics of the building and surrounding. The residual error analysis reveals marked improvements in the main high-error zones, most notably along the Occitanie coast (Figure \ref{fig:dep_costres_agg}). We also observe substantial gains across all performance metrics, with only a moderate increase in the number of parameters (from $19$ to $33$). RMSE decreases from $9000$ to $8781$, and the risk ranking improves, as reflected by the increase in the Gini index (from 0.16 to 0.24).

When the building and local environmental variables are included, several interaction terms become significant. In particular, interactions between rainfall intensity (MILRE) and building characteristics markedly enhance model fit, confirming that flood losses are jointly driven by event severity and local exposure conditions. Among the underwriting variables, the value of mobile assets and the building size remain important, but the interaction between mobile assets and MILRE emerges as the most significant predictor. To better capture exposure mechanisms and mitigate collinearity, we introduce an interaction variable that combines the TRI classification with watercourse-related features (distance to the nearest river and relative altitude difference). This composite indicator, termed \textit{WCTRII}, partitions the TRI zoning into nine exposure categories and proves to be a central predictor in both the severity and occurrence models. While directly incorporating rainfall accumulation into this index did not improve performance, allowing it to interact explicitly with rainfall variables turned out to be more informative.

\begin{figure}[!h]
 \begin{subfigure}{0.49\textwidth}
     \includegraphics[width=0.99\textwidth, height=5cm]{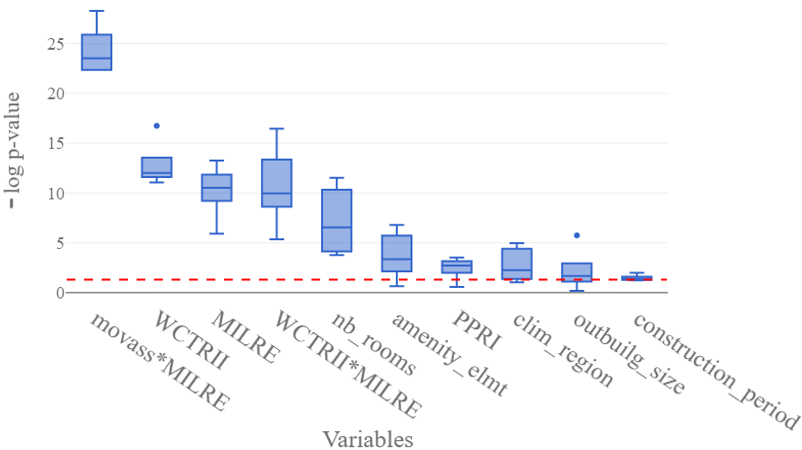}
     \caption{Some cost model variable importance.}
     \label{fig:importance_cost}
 \end{subfigure}
 \hfill
 \begin{subfigure}{0.49\textwidth}
     \includegraphics[width=0.99\textwidth, height=4.8cm]{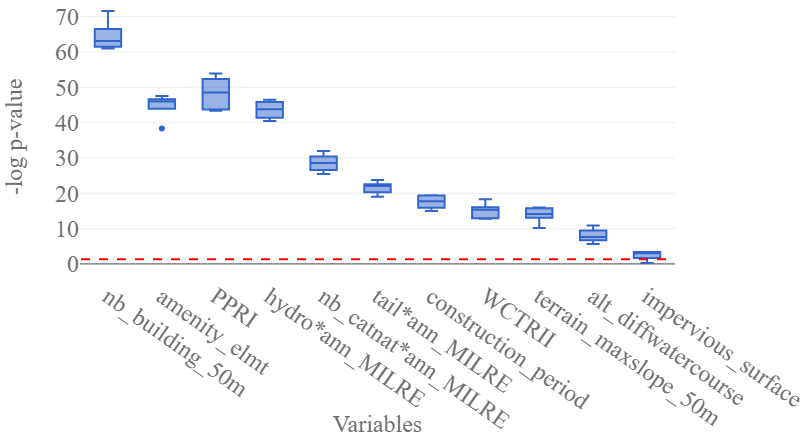}
     \caption{Some probability model variable importance.}
     \label{fig:importance_proba_rank}
 \end{subfigure}
 
 \begin{subfigure}{0.49\textwidth}
     \includegraphics[width=0.99\textwidth, height=5cm]{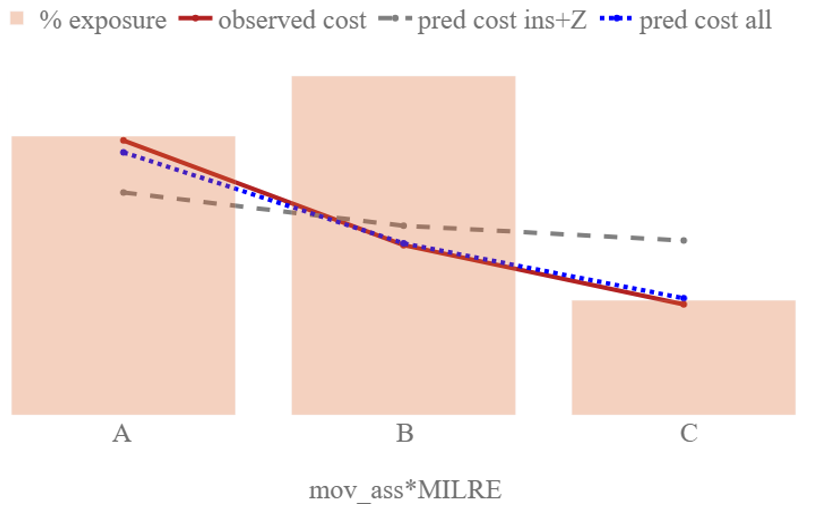}
     \caption{Averaged observed and predicted flood cost regarding \texttt{mov\_ass*MILRE}.}
     \label{fig:movassMILREcost}
 \end{subfigure}
  \hfill
 \begin{subfigure}{0.49\textwidth}
     \includegraphics[width=0.99\textwidth, height=5cm]{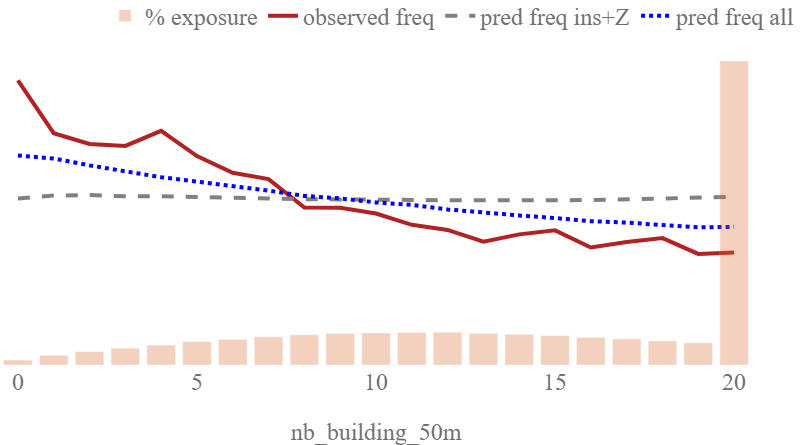}
     \caption{Averaged observed and predicted flood frequency regarding \texttt{nb\_building\_50}.}
     \label{fig:build50proba}
 \end{subfigure}
 \caption{Top plots are the result of a likelihood ratio test (p-value on the log scale) to evaluate variable importance. The dashed red line represent the significance threshold. For the bottom plots, the dataset is ``grouped by'' with regards to the variable, the bars represent total exposure and the dots averaged observed and predicted values by group. Y-axis are masked for confidentiality.}
 \label{fig:importance_cost_and_vars}
\end{figure}

Figure \ref{fig:importance_cost} ranks the main predictive variables using likelihood ratio tests. In general, WCTRII and MILRE dominate the model, followed by building variables provided by insurers that are enriched and supplemented by our dataset. Figure \ref{fig:movassMILREcost} shows how adding the new variables enables us to capture information that was missing from \texttt{ins+c}. Although building-related factors still provide useful signals, their contribution is secondary to that of WCTRII and MILRE, which directly capture flood exposure and event intensity. WCTRII is crucial because it quantifies exposure: high WCTRII scores correspond to properties in low-lying areas close to watercourses, effectively placing them at the forefront of flood risk. MILRE, for its part, characterises rainfall intensity, helping to distinguish routine events from extreme, catastrophic episodes.

\subsection{Probability modelling results and discussions}

\begin{table}[!ht]
\centering
\resizebox{0.9\textwidth}{!}{%
\begin{tabular}{m{3.2cm}m{1.1cm}m{1.1cm}m{1.1cm}m{1.1cm}m{1.1cm}m{1.5cm}}
\toprule
\makecell{Models} & \makecell{-LogLoss} & \makecell{Gini} & \makecell{CSI} & \makecell{Recall} & \makecell{Precision} & \makecell{Nb params} \\ \midrule

 \makecell{\textbf{GLM ins}\\ \emph{\small{insurer data only}}} & \makecell{0.6631} & \makecell{0.2339} & \makecell{0.0049} & \makecell{\textbf{0.1577}} & \makecell{0.0051} & \makecell{\textbf{17}}\\ [0.3cm] 
 
 \makecell{\textbf{GLM ins+c} \\ \emph{\small{insurer data + climate}}} & \formatvarv{0.6245}{\textcolor{blue}{$\shpos5.82\%$}} & \formatvarv{0.3766}{\textcolor{blue}{$\shneg61.01\%$}} & \formatvarv{0.0075}{\textcolor{blue}{$\shneg59.06\%$}} & \formatvarv{0.0902}{\textcolor{red}{$\shpos42.80\%$}} & \formatvarv{0.0083}{\textcolor{blue}{$\shneg62.75\%$}}& \formatvarv{21}{\textcolor{red}{$\shneg23.53\%$}} \\[0.4cm]
 
  \makecell{\textbf{GLM ins+r} \\ \emph{\small{insurer data + rainfall}}} & \formatvarv{0.6082}{\textcolor{blue}{$\shpos8.28\%$}}  & \formatvarv{0.4770}{\textcolor{blue}{$\shneg103.93\%$}} & \formatvarv{0.0159}{\textcolor{blue}{$\shneg224.49\%$}} &\formatvarv{0.0562}{\textcolor{red}{$\shpos64.36\%$}} & \formatvarv{0.0233}{\textcolor{blue}{$\shneg356.86\%$}} & \formatvarv{21}{\textcolor{red}{$\shneg23.53\%$}} \\[0.4cm]
  
   \makecell{\textbf{GLM all} \\ \emph{\small{all data}}} & \formatvarv{\textbf{0.5614}}{\textcolor{blue}{$\shpos15.34\%$}} & \formatvarv{\textbf{0.5462}}{\textcolor{blue}{$\shneg133.52\%$}} & \formatvarv{\textbf{0.0261}}{\textcolor{blue}{$\shneg432.65\%$}} & \formatvarv{0.1280}{\textcolor{red}{$\shpos18.83\%$}}& \formatvarv{\textbf{0.0342}}{\textcolor{blue}{$\shneg570.59\%$}}& \formatvarv{40}{\textcolor{red}{$\shneg135.29\%$}}\\[0.4cm]\bottomrule
\end{tabular}}
\caption{Summary of probability models' metrics calculated with 5-fold cross validation and their relative evolution regarding the insurer data only model. Evolution is in blue when the metric is improving and red if not. Recall and precision are calculated for flooded buildings.}
\label{tab:res_proba_mod}
\end{table}

Table \ref{tab:res_proba_mod} summarises the performance metrics for the probability models. Given the high class imbalance in our dataset, where $99.78\%$ of buildings are non-flooded and only $0.22\%$ are flooded, comparing results to a dummy model that predicts almost exclusively the majority class is not relevant.

\begin{figure}[!h]
     \centering
    \includegraphics[width=\textwidth]{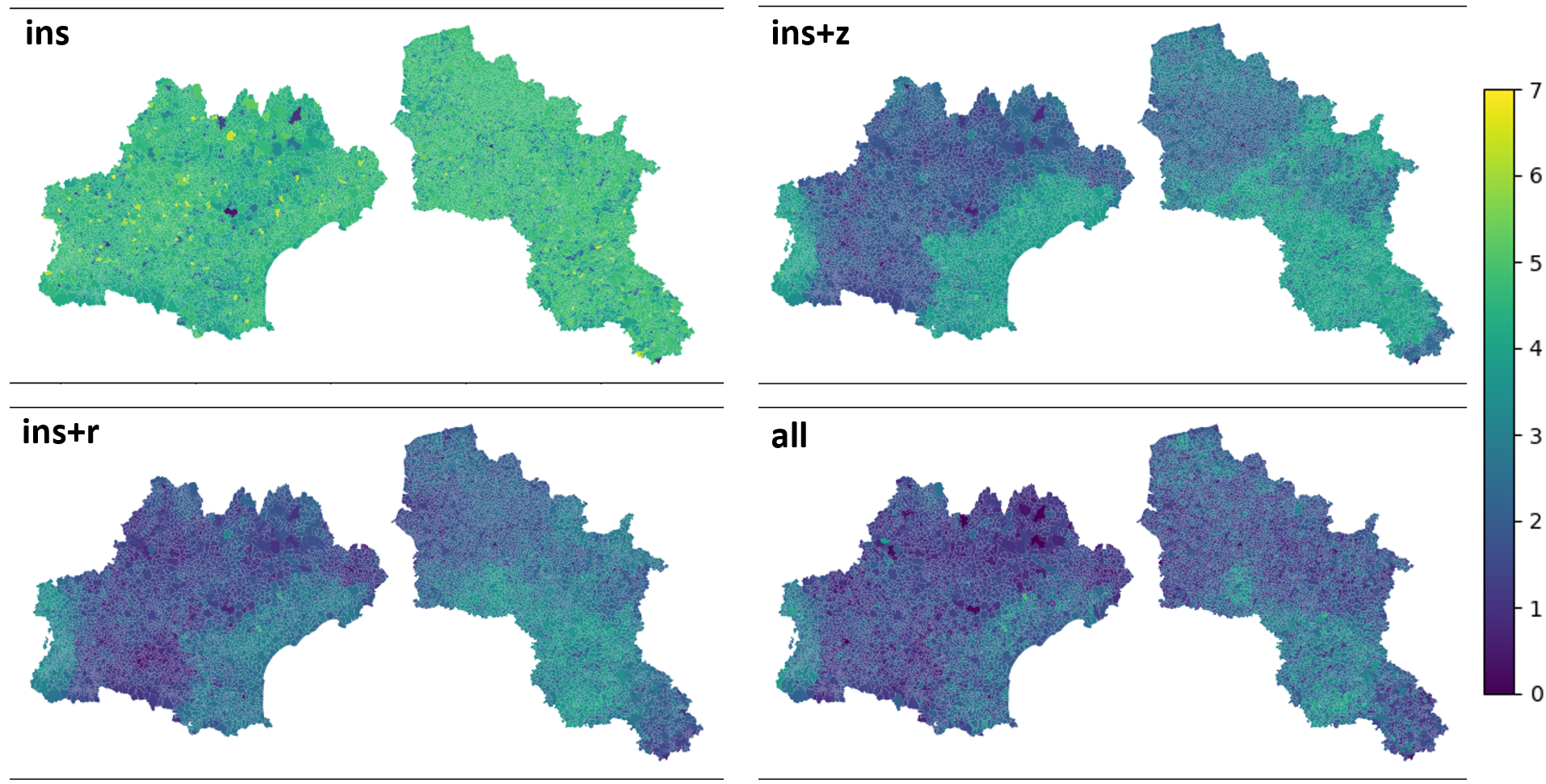}
    \caption{For each of four GLM probability models defined in Table \ref{tab:res_proba_mod} and the two regions defined in Figure \ref{fig:score_tail_cluster1d}, Pearson residuals are displayed at the municipality level. A strong departure from zero indicates a misfit between observed and fitted values.}
    \label{fig:dep_probres_agg}
\end{figure}

\paragraph{\textbf{Insurer model, ``GLM ins'', in table \ref{tab:data_overview} insurance data only}}~~\\ \noindent
As in the cost model, the insurer-only model displays little spatial variation, as shown by the residual maps (Figure \ref{fig:dep_probres_agg}). Under this model, we correctly identify $15.77\%$ of flooded buildings, but this requires labelling $6\%$ of all observations as flooded, resulting in very low CSI and precision. At this point, amenity\_elmt (the presence of pools/outdoor facilities) emerges as the most influential predictor. It reflects differences in vulnerability: outdoor features typically have a lower threshold for damage onset, incurring losses from relatively minor surface runoff events that may not reach or affect the main building structure.

\paragraph{\textbf{Insurer + climate model, ``GLM ins+c'', in table \ref{tab:data_overview} insurance + hydrological and climatic variables}}~~\\ \noindent
Next, we incorporate climate-related expertise variables derived from hydrological flood zones and historical classifications. This extended model shows marked gains in most performance metrics, especially for the Gini and CSI indices (Table \ref{tab:res_proba_mod}), which points to a better probability ranking and calibration.
Under this specification, about $9\%$ of the buildings that actually flooded are correctly detected, while only $2\%$ of the full portfolio is tagged as flooded, reflecting a clear increase in precision. Spatial differentiation improves considerably, although some areas remain overestimated, consistent with the relatively coarse granularity of expert-based zoning (Figure \ref{fig:dep_probres_agg}). A prominent example is southern Occitanie. Due to its history of severe aggregate losses, mainly driven by intense Mediterranean rainfall, the zoning variables uniformly label the entire region as ``high risk'' (see Figure \ref{fig:tri_ppri_interestzone}). Consequently, the model assigns high probabilities throughout this area and does not reflect that actual physical risk is strongly heterogeneous and is largely concentrated along particular watercourses and runoff pathways.

\vspace{0.5cm}
\paragraph{\textbf{Insurer + rainfall model, ``GLM ins+r'', in table \ref{tab:data_overview} insurance + rainfall variables}}~~\\ \noindent
Next, we incorporate information on rainfall intensity via the ann\_MILRE variable, which measures the annual exposure of each building’s location to extreme rainfall episodes. The \texttt{ins+r} model delivers a performance gain, with more precise predictions and improved spatial differentiation. Under this specification, $5\%$ of the flooded buildings are correctly detected, while $0.5\%$ of all the buildings in the sample are flagged as flooded. Overall, the residuals become more structured, notably in high-error zones such as coastal Occitanie and central parts of the northern regions. These enhancements are visible in the residual patterns in Figure \ref{fig:dep_probres_agg}, and are also reflected in higher Gini and precision scores. 

Conceptually, ann\_MILRE acts as a dynamic spatio-temporal filter that efficiently separates year–location combinations exposed to severe precipitation from those facing only limited meteorological risk. In the data, floods almost never occur when ann\_MILRE remains below moderate levels. This indicates that very intense rainfall is a necessary, though not sufficient, condition for flooding. As a result, we interpret this variable as guiding the model to ``concentrate'' its learning on more difficult-to-classify areas where a potential flood-triggering event has taken place.

\paragraph{\textbf{All variables model, ``GLM all'', in table \ref{tab:data_overview} insurance + rainfall + climate + building variables}} \noindent
Finally, we develop a comprehensive model that combines all available data sources. In this specification, we also incorporate the WCTRII variable, which was significant in the cost model and remains relevant for probability estimation. This model yields a substantial gain in probability calibration, of magnitude comparable to the initial improvement achieved when spatial information was first introduced (see metrics in Table \ref{tab:res_proba_mod}). It correctly detects $12.80\%$ of flooded buildings while flagging only $0.7\%$ of the portfolio as flooded, further limiting false positives. The precision of predictions for flooded buildings increases considerably, while a high recall level is maintained unlike in the preceding model. The residual map appears darker, reflecting improved predictive performance across all regions. In addition, high-risk areas that were previously overstated due to broad-scale patterns in open data are now delineated more accurately, highlighting one of the key advantages of high-resolution geolocated information. Compared with coarse zoning classifications, such data support a much finer risk stratification, lowering the chances that large segments of the portfolio are erroneously tagged as high risk or overlooked as low risk.

Figure \ref{fig:importance_proba_rank} shows that building attributes and environmental context exert a stronger influence in the probability model than in the cost model. The most important predictor is nb\_building\_50m, followed closely by amenity\_elmt. The variable nb\_building\_50m, which quantifies the number of buildings within a 50 m radius, is particularly informative in the probability model because it captures both urbanisation patterns and spatial constraints imposed by nearby water bodies. Areas with high flood risk typically feature lower building densities, shaped by zoning regulations and the legacy of past floods. As a result, buildings in these zones tend to have low nb\_building\_50m values, making this indicator a useful spatial proxy for flood exposure. When a building is located near a river or other water body, part of the 50 m buffer overlaps with water, reducing the number of surrounding buildings counted. This natural truncation effectively reflects the heightened flood risk associated with close proximity to water, as such buildings are more exposed to overflow. Furthermore, topography and flood protection measures influence building density, likely reinforcing the link between low nb\_building\_50m values and flood risk.

The variable ann\_MILRE appears in three of the top interaction terms, confirming its key role in capturing annual variability in risk (Figure \ref{fig:importance_proba_rank}). The variable nb\_catnat flags areas with a history of flooding, whether frequent, occasional, or rare. Although informative, this is static information: in a year without an extreme event, even a historically flood-prone area may present little actual risk, whereas a zone with few past floods can be heavily impacted in a year with severe events. This is where interactions with ann\_MILRE become valuable, as they contextualise historical information, offering a dynamic risk assessment.

\subsection{Discussions}
Official flood zoning is a key component of flood risk management in France. Among the available datasets, the \textit{Territoires à Risque Important d’Inondation} (TRI) provide the most detailed and authoritative delineation of flood-prone areas, even though their spatial coverage is restricted (mainly due to implementation costs) and they are designed primarily for regulatory planning rather than insurance pricing. In what follows, we therefore use TRI both as an interpretive benchmark and as a reference ground truth.

Estimates from the cost and probability models show that, where available, official zoning has strong explanatory power. Only $5\%$ of buildings in our data are within the TRI flood zones, yet these zones account for about $11\%$ of recorded flood events. The contrast increases in high-risk TRI zones, which contain under $1\%$ of buildings but roughly $4\%$ of flood occurrences, with average costs 1.6 times higher than for buildings outside TRI. This concentration supports the relevance of expert-based flood mapping for pinpointing structurally exposed areas.
Within TRI zones, models that include zoning data and the fully integrated specification behave almost identically, both in predicted probabilities and loss levels. This similarity suggests that the extra variables in our framework align with, rather than challenge, the expert judgment embedded in regulatory flood maps. The framework’s main added value arises outside TRI-designated areas, which make up about $95\%$ of the portfolio. In these regions, official flood maps offer little to no guidance, so insurers must turn to other information sources to distinguish risk levels.

In the cost model, combining rainfall intensity (MILRE) with detailed environmental indicators and watercourse-related variables makes it possible to isolate groups of non-TRI buildings with significantly higher average losses, a difference that is essential for pricing. In these high-risk groups outside the TRI zones, the insurer's baseline model plus climate variables explains only 57\% of the total observed costs, compared to 97\% for the fully integrated model. However, across the entire non-TRI area, both models explain a similar proportion of total losses. This shows that the extra variables do not raise overall risk estimates but instead sharpen how losses are distributed within unzoned areas, cutting underpricing for highly exposed profiles and curbing overpricing for lower-risk ones.

A comparable trend emerges in the probability models. Outside TRI zones, the \texttt{ins+c} model systematically overestimates the probability for non-flooded buildings, assigning values about $15\%$ higher than those of the \texttt{all} model. In contrast, for buildings that have been flooded, the \texttt{all} model produces probabilities roughly $10\%$ higher on average, indicating better calibration and more accurate detection of vulnerable structures. In particular, the \texttt{all} model also highlights high-risk zones absent from official climate risk inventories (TRI, PPRI, etc.) but characterised by clear aggravating factors such as steep slopes, natural retention areas, large impervious surfaces and heavy-tailed rainfall. We interpret this as flagging locations where official flood maps may be incomplete, outdated, or not configured to capture certain flood processes. The model further identifies lesser-known waterways that may pose hazards despite not being classified as major flood-prone areas.

The municipality of Le Perray-en-Yvelines and its surroundings, south of Paris, offer a telling case study. In recent years, upstream urbanisation has markedly expanded impervious surfaces, modifying local hydrological behaviour during heavy rainfall. Although runoff-induced flooding has been mentioned in local reports, the area was outside the TRI and PPRI boundaries, and no floods were listed in the GASPAR database during our calibration period. Despite this lack of official zoning, the \texttt{all} model identifies several buildings in the municipality as high risk, in contrast to the climate-adjusted insurer model (\texttt{ins+c}), which rates them close to the portfolio average. This divergence arises because the \texttt{all} model accounts for imperviousness, terrain slope, and rainfall extremes. During this study, in November 2024, Storm Kirk brought flooding to Perray-en-Yvelines. Two subsequent ministerial decrees officially recognised flood events there, the first ever for the municipality. Although this is a single case, this episode shows how insurance-based models that incorporate fine-scale exposure and rainfall information can highlight emerging flood risks before they appear in official datasets.

\begin{figure}[!h]
     \centering
    \includegraphics[width=\textwidth]{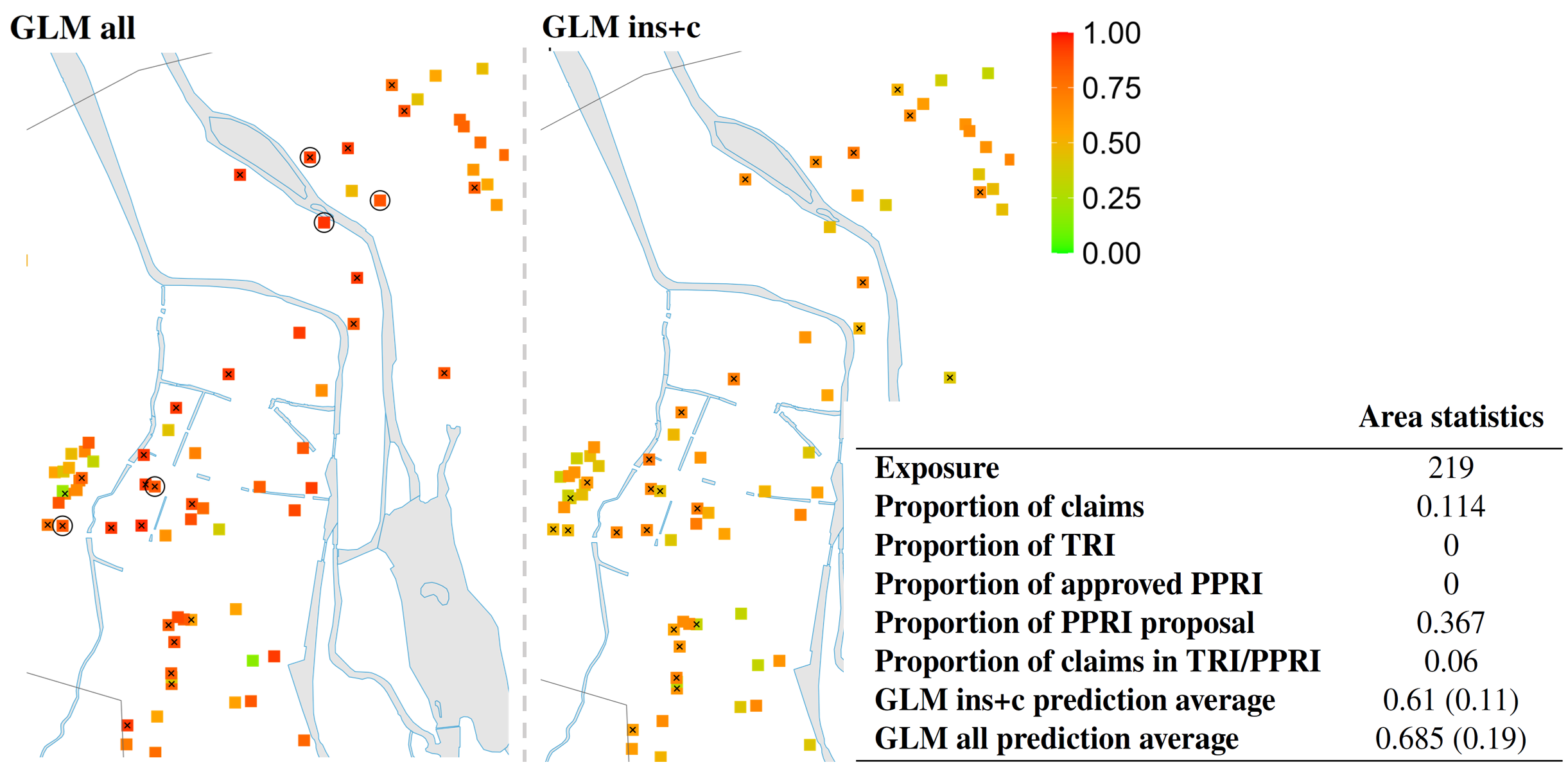}
    \caption{Comparison of predicted flood occurrence probabilities from the \texttt{ins+c} and \texttt{all} models for a segment of the city of Montargis. Each square represents a building; crosses indicate buildings with at least one observed flood claim in the portfolio. Circled buildings correspond to locations where the difference between model predictions is largest. The table summarises key statistics for the displayed area.}
    \label{fig:example_1_montargis}
\end{figure}
\vspace{3mm}
The city of Montargis (Figure~\ref{fig:example_1_montargis}) illustrates a situation where flood risk is well known but only partially formalised. The municipality has faced repeated flooding, with twelve ministerial flood recognitions listed in the GASPAR database since 1980. However, no TRI has been designated and only an unapproved PPRI has existed since 2018, likely reflecting the modest population of the town (15,000 inhabitants) and limited economic stakes. In this setting, the climate-augmented insurer model (\texttt{ins+c}) generates broadly uniform probability estimates over much of the urban area, mainly informed by coarse-scale climatic and zoning variables. In contrast, the \texttt{all} model adds finer spatial details by including indicators such as distance to rivers, relative elevation, and local building layout. As shown in Figure~\ref{fig:example_1_montargis}, this yields higher predicted probabilities along river corridors and in low-lying zones, where claims are also more common, while leaving the rest of the municipality closer to baseline risk. The circled buildings mark the locations where the two models diverge the most, revealing exposure patterns that merit closer examination.

\begin{figure}[!h]
     \centering
    \includegraphics[width=\textwidth]{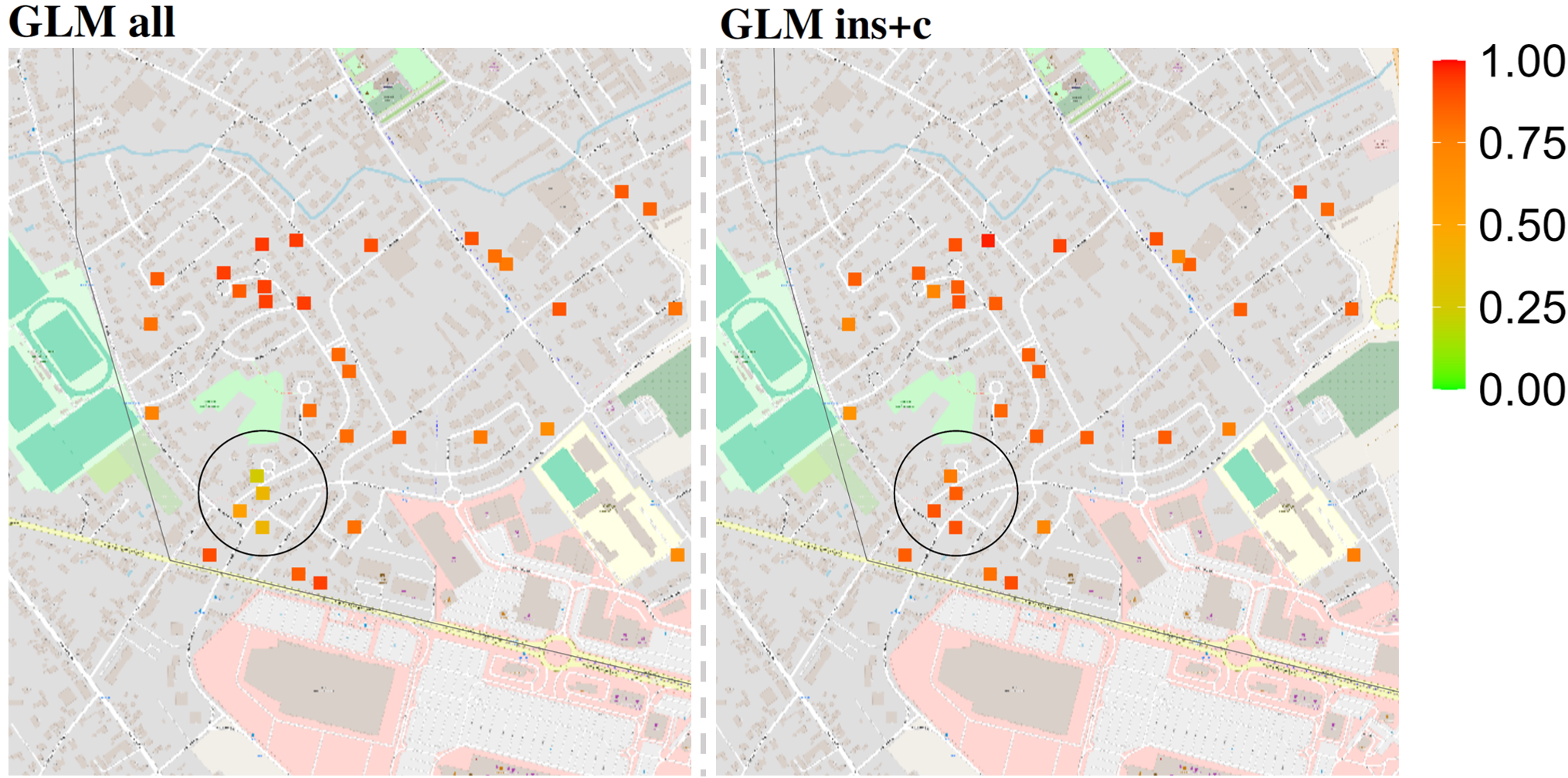}
    \caption{Comparison of predicted flood occurrence probabilities from the \texttt{ins+c} and \texttt{all} models for a segment of the city of Saint-Memmie. Each square represents a building. Circled areas highlight locations where the two models yield the most divergent probability estimates.}
    \label{fig:example_2_saintmemmie}
\end{figure}

\vspace{3mm}
Figure~\ref{fig:example_2_saintmemmie} shows a contrasting situation for Saint-Memmie, where there is both an approved PPRI and officially recognised flood events. In this context, the \texttt{ins+c} model already captures much of the main spatial pattern of flood risk, producing higher than average but fairly homogeneous predictions within the regulated floodplain. The added value of the \texttt{all} model lies in refining risk estimates locally. The circled area in Figure~\ref{fig:example_2_saintmemmie} contains buildings higher than the river, with limited slope-driven runoff and a lower propensity to water accumulation. For these buildings, the \texttt{all} model assigns smaller probabilities than \texttt{ins+c}, indicating reduced exposure despite their location within officially mapped large-scale flood zones. Recall that PPRI coverage is extensive, but relatively coarse. This case underlines how fine-grained data can complement existing zoning: when regulatory maps are available, high-resolution environmental and building attributes help prevent over-smoothing by revealing micro-scale exposure contrasts inside broadly defined risk areas.

\section{Conclusion}\label{sec:conclusion}

Using a conventional GLM framework to isolate the contribution of data rather than modelling techniques, this study examines how the integration of detailed geolocated information improves the assessment of flood risk for insurance. Models relying solely on underwriting data exhibit weak explanatory and spatial discriminatory power, while the inclusion of official zoning and expert climate and hydrographic variables significantly enhances performance where such information is available. However, the greatest gains come from incorporating high-resolution environmental and rainfall data at the building level, which better capture local exposure mechanisms and event intensity. This refined modelling is especially valuable outside official flood zones, allowing more granular risk allocation, more accurate pricing of highly exposed properties, and reduced overpricing of lower-risk ones, without challenging existing regulatory classifications.

The role of explanatory variables differs by modelling objective. Severity models are primarily driven by indicators of exposure and event intensity (e.g. proximity to watercourses, relative elevation, extreme rainfall metrics), complemented by insured values that scale losses once flooding occurs. Other added building variables thus play a limited role. By contrast, occurrence models rely more on descriptors of the immediate environment such as building density, land cover, and local topography, which govern whether flooding occurs at all. They also gain from annual aggregated rainfall indicators. 

In this study, rainfall variables are not designed for short-term or real-time forecasting, which is beyond our scope. Instead, for pricing and portfolio risk assessment, they can be incorporated by conditioning expected losses on different intensity levels and integrating over their empirical distribution, yielding unconditional risk estimates consistent with actuarial principles. Beyond their direct predictive contribution, these variables stabilise the models by capturing event severity, reducing omitted-variable bias, and improving the robustness and interpretability of other covariate predicted effects. The aggregation choices reflect insurance use cases and aim to capture the most intense rainfall events across multiple time horizons. During model development, alternative aggregation windows were tested and produced consistent results, indicating that the key factor is the ability to contrast short- and long-term intensity measures rather than the exact temporal definition. Although calibrated on French data, the rainfall metric is based on general principles of local extreme precipitation exposure and is therefore potentially transferable to other regions with comparable rainfall information.

Beyond pricing, these results have implications for prevention and adaptation. By highlighting the influence of local small-scale features, insurance-based models can complement public hazard maps and support more targeted risk awareness and mitigation efforts. In this sense, high-resolution data should not be viewed as a tool for exclusion, but as a means to strengthen collective resilience through better-informed decisions. In the French CatNat system, which is based on national solidarity, improved risk differentiation is not intended to restrict coverage. Rather, more precise modelling can support fairer pricing, improved portfolio monitoring, and more effective prevention strategies, while remaining consistent with existing regulatory frameworks.

Overall, this study shows that actuarial models enriched with geolocated environmental data can usefully complement official flood risk assessments by adding spatial detail and insurance-relevant perspectives on exposure. More broadly, enhancing flood risk assessment through data integration is a key component of long-term climate resilience and preparedness.

\paragraph{Acknowledgments}
We are grateful to ADDACTIS France for providing access to the insurance data used in this study.


\paragraph{Competing Interests}
None.

\paragraph{Data Availability Statement}
All data used in this study to build explanatory variables are open source. The french disaster database (GASPAR) is also open source. Due to confidentiality reasons, the insurance portfolio used in this study is not available. 

\paragraph{Funding statement}
This work received no specific grant from any funding agency, commercial or not-for-profit sectors.





\printendnotes
\printbibliography

\appendix

\section{Complementary details on data sources and variable construction}\label{appendixA}

\subsection{Additional information on hydrological and climatic indicators} 

\begin{itemize}
\item \textbf{TRI (Territoires à Risque d’Inondation):} TRI zones were introduced under the European Floods Directive (2007/60/EC) to pinpoint areas subject to significant flood risk and to guide the prioritisation of flood risk management measures at both national and European scales (\cite{CEPRI2022floodlaw}). Their purpose is to identify territories where high flood hazard coincides with dense populations, substantial assets, or intense economic activity, rather than to produce comprehensive flood maps for the whole territory.

The definition of TRI zones relies on a combination of data sources, including documented past flood events, hydrological and hydraulic modelling of representative flood scenarios, and exposure indicators such as population density and economic value. These components are analysed together to evaluate the potential flood impacts at a territorial level. Consequently, TRI boundaries embody a synthetic assessment of flood risk that incorporates both hazard and exposure dimensions.
Within TRI boundaries, flood hazards are grouped into high, medium, and low categories, generally corresponding to decreasing exceedance probabilities, often interpreted as indicative return periods of about 10, 100, and 1,000 years. These categories characterize the anticipated intensity and spatial extent of flooding under reference scenarios. They do not explicitly account for building-level vulnerability, local protection measures (in some instances), or year-to-year variability in flood occurrence. It should be emphasized that the methodology for constructing TRI is tightly regulated, yet it can differ from one region to another. The availability of high-quality hydraulic modelling, advanced digital elevation models, and other data can affect the overall quality of a given TRI. Despite these limitations, TRI currently represent the most advanced level of flood assessment produced by French public institutions.

By design, TRI coverage is restricted to areas deemed nationally significant in terms of flood risk. As a result, a substantial portion of French territory is not classified. In our dataset, around $95\%$ of buildings are located outside TRI perimeters. We therefore introduce an explicit ``none'' category to distinguish areas that lie beyond TRI coverage from those that are classified as low hazard within TRI zones.

\item \textbf{PPRI (Plans de Prévention des Risques Naturels d’Inondation):} these maps are local regulatory tools established at the municipal or inter-municipal scale to guide land-use planning and limit exposure to flooding (\cite{PPRI2023}). In contrast to TRI, which are intended to support national prioritisation, PPRI are designed for local decision-making and zoning regulations.

PPRI are generally derived from hydraulic simulations of a reference flood, most commonly defined by a 100-year return period, and can be supplemented with historical extreme events when these exceed the reference scenario. The simulations use river geometry, topography from digital elevation models, and simplified flow representations, but typically do not incorporate prospective land-use changes, climate evolution, or building-scale mitigation actions. The resulting maps classify areas into hazard categories ranging from very low to very high, which are then converted into regulatory constraints on land use. In our dataset, detailed intra-municipal PPRI represent $1.4\%$ of the portfolio, with around $20\%$ overlapping with TRI zones.

PPRI coverage is uneven because their development and approval demand substantial technical and administrative resources. As a result, many municipalities do not have detailed intra-municipal mapping and rely solely on a municipality-wide evaluation of potential flood risk, which may be approved, outdated, or still under review (unapproved). Regarding TRI, we add a “none” category to explicitly indicate when PPRI information is missing. From a modelling standpoint, detailed PPRI serve a function similar to TRI, whereas municipality-level PPRI extend spatial coverage and enhance model performance, but at the expense of spatial accuracy and by relying on a coarser classification of potentially exposed areas.

Figure \ref{fig:tri_ppri_interestzone} provides an overview of flood-prone areas in our study regions, as defined by TRI and PPRI, highlighting zones with known exposure.

\begin{figure}
 \begin{subfigure}{0.49\textwidth}
     \includegraphics[width=0.9\textwidth]{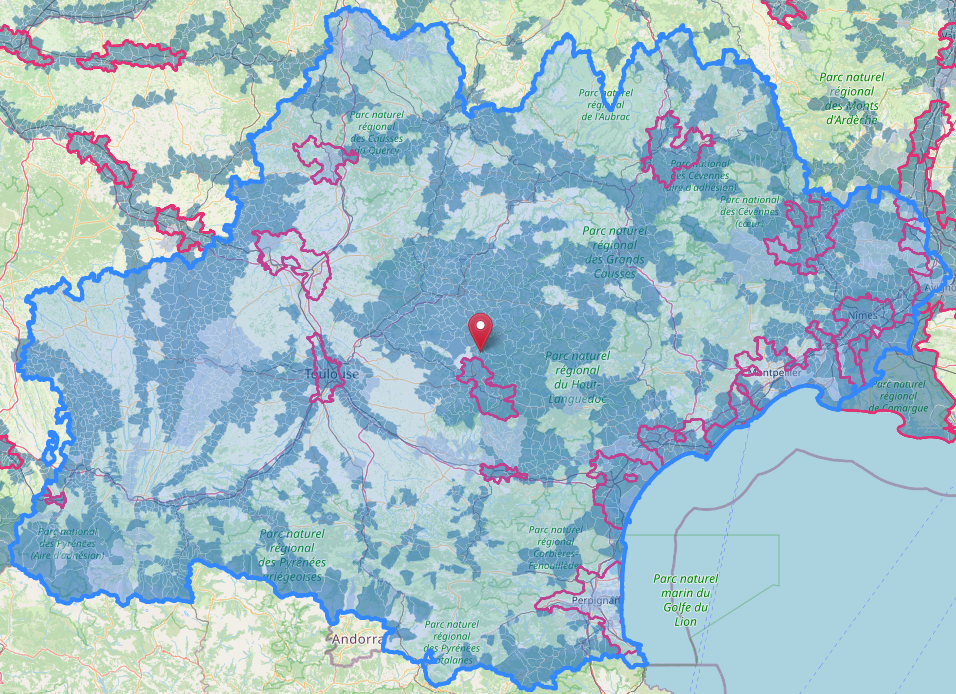}
     \caption{Occitanie}
     \label{fig:occ_tri_ppri}
 \end{subfigure}
 \hfill
 \begin{subfigure}{0.49\textwidth}
     \includegraphics[width=0.9\textwidth]{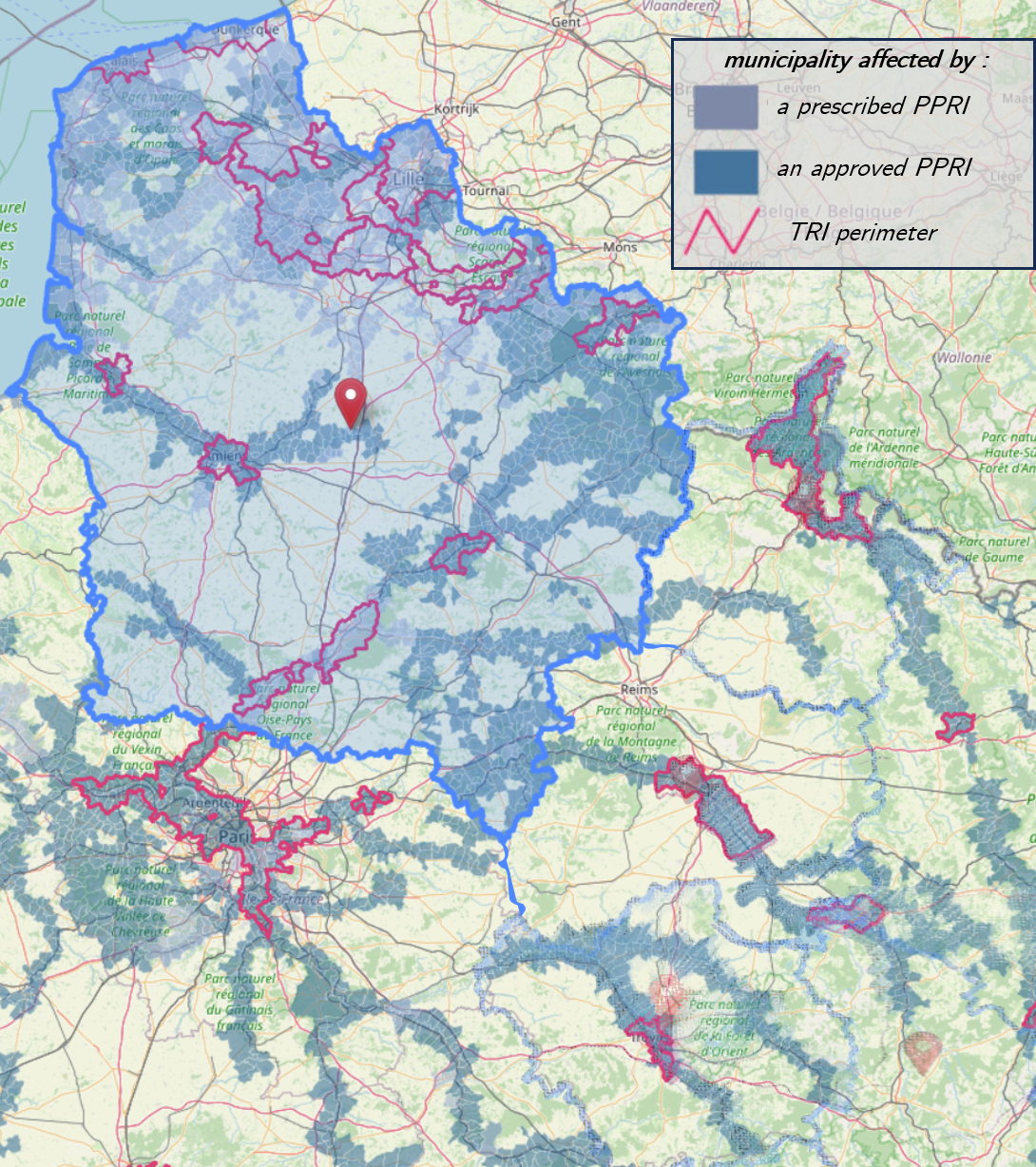}
     \caption{Haut-de-France and Grand-Est westside}
     \label{fig:hdf_tri_ppri}
 \end{subfigure}

 \caption{French expert reference maps for flood risks assessment. PPRI (Plans de Prévention Risque naturelle Inondation) represent zones exposed to flooding from 100-year reference event, supporting land-use planning  and risk mitigation. PPRI are approved by the departement's prefect. TRI (Territoire à Risque Inondation) are elaborated for high risk flooding zones with high stakes. Three scenario (low, medium and high return period) are used to define exposed surface and associated risks.}
 \label{fig:tri_ppri_interestzone}
\end{figure}

\item \textbf{Hydrographic sectors:} the French hydrographic system is structured hierarchically into four nested levels: hydrographic regions, sectors, sub-sectors, and zones, corresponding to successive subdivisions of the main river basins defined under the Water Framework Directive (\cite{EU2000directive}). These basins are defined as areas in which all surface runoff converges toward a common outlet through a connected river network, making them a natural spatial reference for fluvial flood processes.

In this study, we use the hydrographic sector level, which partitions mainland France into 24 units characterised by distinct river systems, watershed structures, and hydrological responses (\cite{regionhydro2019}). This variable is not intended to model upstream–downstream accumulation or dynamic catchment processes explicitly. Instead, it serves as a regional segmentation factor capturing broad differences in hydrological behaviour, such as runoff generation, river density, and dominant flood mechanisms. Similar hydrographic divisions are commonly used in the literature to control for spatially coherent hydrological heterogeneity (\cite{fortin2020risques}).

\item \textbf{Climatic regions:} climatic zoning reflects long-term spatial patterns in precipitation, temperature, and atmospheric circulation that shape flood-generating processes. Several climatological typologies identify distinct climate regimes across mainland France (\cite{MF-decoupmet,joly2010types}). A commonly adopted classification distinguishes five major climate types: oceanic, altered oceanic, continental, Mediterranean, and mountain climates, each associated with specific rainfall seasonality and extreme precipitation characteristics.

Météo-France further refines this framework into 29 climatic sub-regions to capture finer spatial variability in rainfall intensity, storm typology, and hydrometeorological dynamics. We adopt this refined zoning as a contextual variable to account for regional differences in precipitation regimes and flood exposure that are not explicitly modelled at the event scale. As with hydrographic sectors, climatic regions act as structural controls rather than predictors of individual flood events.

\item \textbf{GASPAR database:} this database (Base nationale de Gestion Assistée des Procédures Administratives relatives aux Risques) is a national administrative registry that inventories official recognitions of natural disasters under the French CatNat insurance scheme (\cite{gaspar2023}). For flooding, it records the affected municipalities, dates, and hazard types associated with each ministerial decree.

This dataset provides a valuable historical reference on the spatial and temporal distribution of major flood events. However, inclusion requires a formal request by municipalities and approval by an interministerial committee, meaning that GASPAR primarily captures events deemed significant at the administrative level. As a result, localised, low-intensity, or unreported floods may be underrepresented. In this study, GASPAR is therefore used as a contextual indicator of historical flood occurrence rather than as an exhaustive catalogue of flood events.

\end{itemize}

\subsection{From an address to variables} 

As a general rule, insurance datasets do not provide the exact location of insured buildings; instead, they usually contain only the municipality code. Although this level of detail is adequate for most spatial analyses and zoning-based risk evaluations (\cite{wahl2022spatial}), accurate geolocation allows the inclusion of fine-scale environmental and structural information. To leverage this potential, we design a multi-step geocoding workflow that links policyholder addresses to individual buildings. We first clean and standardise address records so that they are compatible with the Base Adresse Nationale (BAN), the official French address database. This harmonisation improves consistency and improves matching performance. Each normalised address is then associated with one or more buildings drawn from a consolidated building database assembled from several open data sources, covering all known geolocated buildings in the study region. In $25\%$ of instances, an address corresponds to multiple buildings; in these cases, we retain only the principal building, as the remaining buildings are typically outbuildings or minor structures.  
To reduce misclassification in building assignment, we apply integrity checks based on business rules and insurer-supplied information, ensuring alignment with policyholder records. After the insured building and its precise GPS coordinates have been determined, a combination of machine learning, computer vision, Natural Language Processing, and Geographic Information System techniques can be employed to automatically extract information from maps, imagery, textual documents, and related sources. This enables the construction of detailed variables that describe both the building and its surrounding environment.
For a more detailed description of the methodology employed to convert raw data into structured risk variables, along with illustrative examples, see \cite{chatelain2023tarification}. This reference discusses the development of a building and surroundings dataset and demonstrates its use in insurance risk modeling.

\subsection{Building and surrounding data construction methods}\label{appendixA1}

Depending on the structure and constraints of the underlying data sources, different processing strategies are applied to build the explanatory variables, which are described in the following sections. Illustrative results are shown at the municipal level, where the median is used to summarise the behaviour of each variable in the study area. \vspace{2mm}

\begin{figure}
 \begin{subfigure}{0.49\textwidth}
     \includegraphics[width=1\textwidth, height=4.5cm]{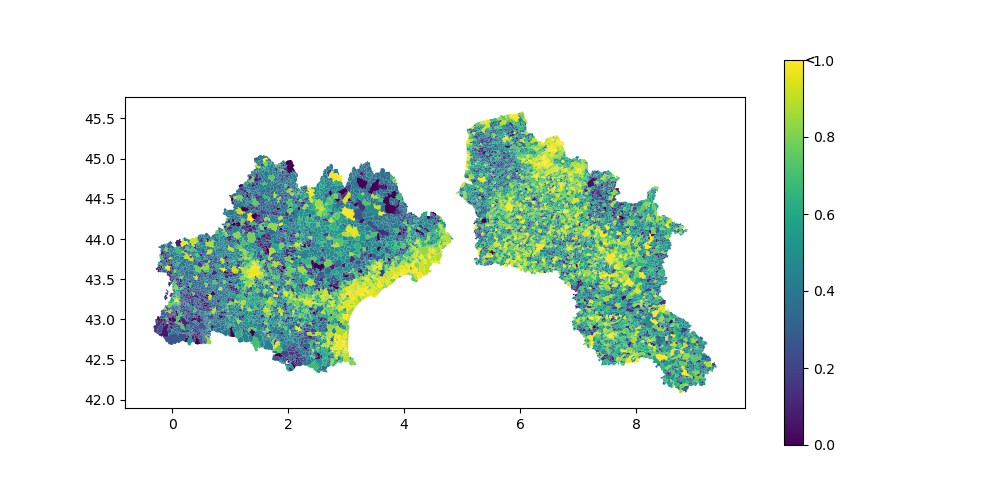}
    \caption{Percentage of impervious surface in a 200m buffer around the building, example of integrated data.}
    \label{fig:impervioussurface}
 \end{subfigure}
 \hfill
 \begin{subfigure}{0.49\textwidth}
     \includegraphics[width=1\textwidth, height=4.5cm]{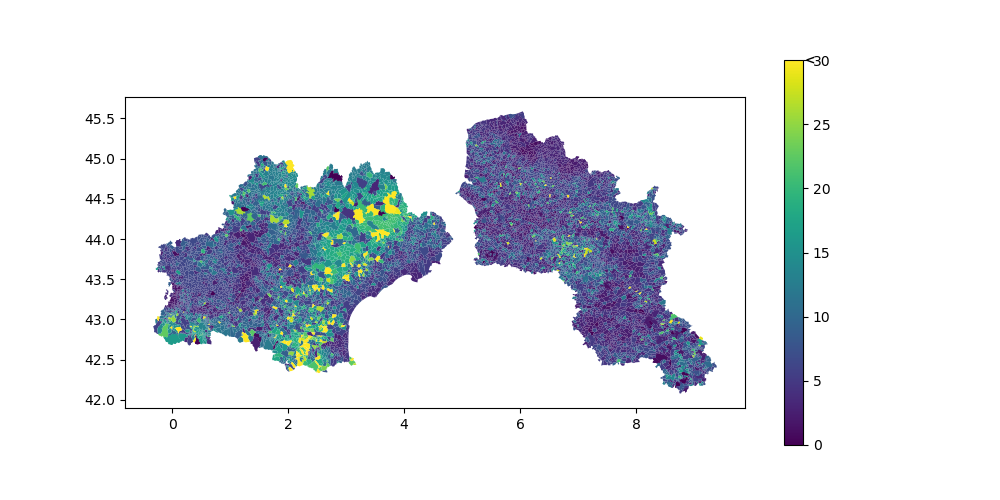}
    \caption{Highest altitude difference in meters between a given building and the environment in a 100m buffer, example of computed data.}
    \label{fig:slope100m}
 \end{subfigure}
 \caption{Variables aggregated by median at municipality level, illustrating four different approaches to data construction.}
 \label{fig:ex_2}
\end{figure}

\textbf{Integrated data} correspond to structured open datasets that can be directly linked to the insurance portfolio with limited transformation. Depending on their spatial nature, integration relies either on administrative identifiers (e.g. municipal codes) or on spatial joins between geometries. Figure \ref{fig:impervioussurface} illustrates the proportion of impervious surfaces within a 200-metre buffer around each building, derived from the national land-cover dataset ``Occupation des Sols'' (\cite{occupsols2023}). This raster-based product, available at 10-metre resolution, is intersected with building neighbourhoods to quantify local soil sealing. Higher values are observed in and around major urban areas, with a clear spatial gradient across regions.

In contrast to completed, created, and computed variables, integrated data do not rely on inference or modelling. The subsequent data categories represent more innovative features, either because they are derived from unstructured sources, require substantial processing to be exploitable at the building scale, or combine multiple contextual inputs that are not jointly available in standard datasets.

\vspace{2mm} 

\begin{figure}
 \begin{subfigure}{0.49\textwidth}
     \includegraphics[width=1\textwidth, height=4.5cm]{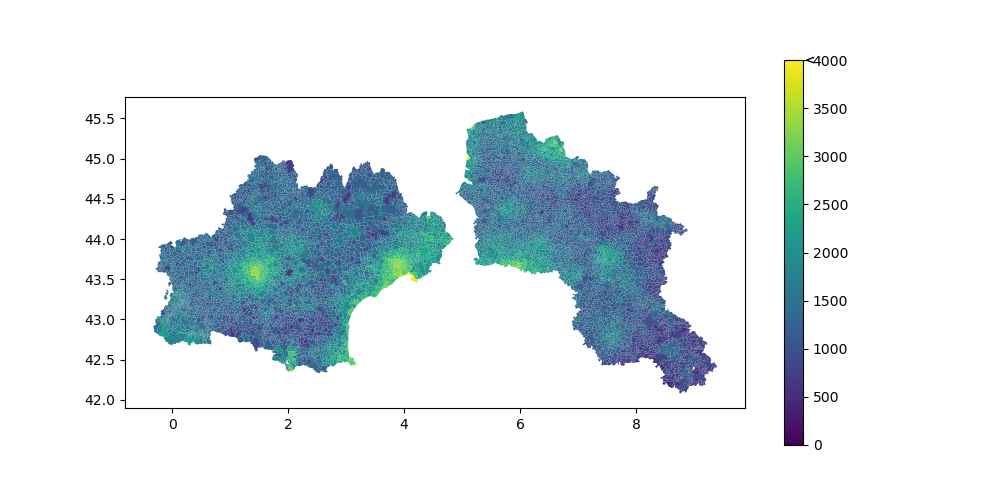}
     \caption{Property value, example of completed data.}
     \label{fig:housevalue}
 \end{subfigure}
 \hfill
 \begin{subfigure}{0.49\textwidth}
     \includegraphics[width=1\textwidth, height=4.5cm]{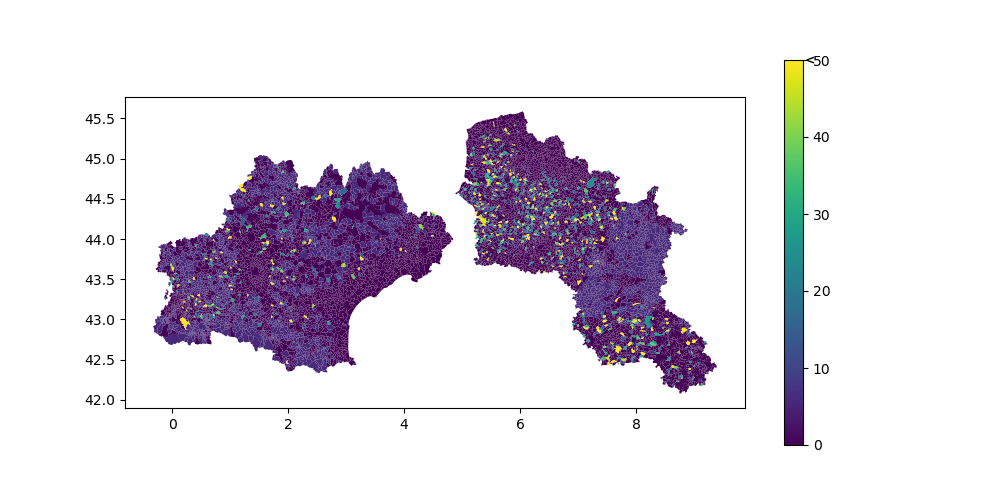}
     \caption{Annex building presence, example of created data.}
     \label{fig:annexbuild}
 \end{subfigure}
 \caption{Variables aggregated by median at municipality level.}
 \label{fig:ex_1}
\end{figure}

\textbf{Completed data} originate from open or administrative data sources that are exhaustive in scope but incomplete at the building level, limiting their direct usability for insurance modelling. Missing values are inferred using supervised machine learning models trained on well-documented observations and enriched with internal insurance data, building characteristics, and spatial context. A typical example is the value of the property. Transaction prices are available through the DVF database, which records all real-estate transactions in France (\cite{dvf2019}), but only a small fraction of buildings have been sold in recent years. To obtain a complete building-level variable, predictive models are trained on observed prices using explanatory features such as the footprint of the building, the number of floors, the location, and the surrounding characteristics, and then used to impute missing values. Figure \ref{fig:housevalue} illustrates the resulting average house values, with higher levels concentrated in major urban areas.

\vspace{3mm} 
\textbf{Created data} are derived using advanced data-processing techniques, including computer vision and natural language processing, applied to raw, unstructured sources such as aerial imagery or administrative text. These variables capture information that is not directly available in conventional datasets, for example, the presence and size of secondary outbuildings. To ensure reliability, generated variables are systematically validated against manually labelled reference samples, comparing predicted outputs with pixel-level image annotations or curated textual extracts. Figure \ref{fig:annexbuild} illustrates the spatial distribution of secondary outbuildings, showing a lower prevalence in dense urban areas. This variable exemplifies created data, as it is produced by applying image-recognition models to satellite imagery rather than by completing an existing structured dataset.

\vspace{2mm} 
\textbf{Computed data} Computed variables are derived through the combination of multiple datasets using spatial analysis or simplified physical metrics. They typically require geometric operations, neighbourhood aggregation, or the transformation of continuous spatial fields into building-level indicators. An illustrative example is shown in Figure \ref{fig:slope100m}, which represents the maximum elevation difference between each building and its surrounding terrain within a radius of 100 metres. This variable is computed using the national digital elevation model (\cite{alti2022}) by spatially intersecting elevation data with building footprints and summarising local altitude gradients. It captures local topographic contrasts that influence runoff concentration and flow direction. As expected, higher values are observed in mountainous regions such as Occitanie.

\newgeometry{left=2cm,bottom=0.1cm,top=0.5cm}  

\newcommand{\formatvar}[2]{\makecell[l]{\texttt{#1}\\\emph{\small{#2}}}}
\newcommand{\formatres}{l}
\newcommand{\bob}{l}

\begin{table}[!ht]
\centering
\resizebox{\textwidth}{!}{%
\begin{tabular}{p{4.4cm}p{2.1cm}p{2.9cm}p{2.9cm}p{5.2cm}}

\toprule
\textbf{Variable name}  & \textbf{Value taken} & \textbf{global stats} & \textbf{flooded stats} & \textbf{Details} \\ [0.1cm] \midrule
\textbf{\underline{Insurance}}&  & \makecell[l]{produced on all\\ buildings}  & \makecell[l]{produced on flooded \\buildings}  &  stats values: \makecell[l]{average (standard deviation)\\ or proportion \%}   \\ [0.3cm]

\formatvar{claim\_amount}{individual claim expenses}& $[5,69k]$ & - & $5884.13~~(9093.98)$ & \makecell[\bob{}]{claim amount for a given contract\\ and flood occurrence} \\ [0.3cm]

\formatvar{claim\_nb}{number of claims}   & $\{0,1\}$   & $0.221\%~~(0.0473)$  & - & -\\ [0.3cm]

\formatvar{nb\_rooms}{number of rooms}& $[2,6]$  & $3.68~~(0.8)$ &  $3.86~~(0.79)$& -   \\ [0.3cm]


\formatvar{mov\_assets}{movable assets value}& $[0,456k]$  & \makecell[\formatres{}]{$[5k,10k]:15.77\%$ \\$[10k,20k]:14.06\%$}  & \makecell[\formatres{}]{$[30k,45k]:16.16\%$ \\$[10k,20k]:14.82\%$} & continuous variable used in a discretized for confidentiality purpose \\ [0.5cm]

\formatvar{prec\_obj}{precious object value}& $[0,25k]$ & \makecell[\formatres{}]{$[0,5k]:91.34\%$ \\$[5k,+\infty]:8.65\%$} & \makecell[\formatres{}]{$[0,5k]:86.91\%$ \\$[5k,+\infty]:13.09\%$} & continuous variable used in a discretized for confidentiality purpose \\ [0.5cm]

\formatvar{amenity\_elmt}{amenity elements}&  {pres, abs} & \makecell[\formatres{}]{abs: $93.01\%$ \\ pres: $6.99$} & \makecell[\formatres{}]{$abs:84.18\%$ \\ $pres:15.82$} & principally presence of pools   \\ [0.5cm]

\formatvar{outbuilg\_size}{eventual outbuildings size}& {pres, abs} & \makecell[\formatres{}]{abs: $98.56\%$ \\ pres: $1.44$}  & \makecell[\formatres{}]{$abs:98.25\%$ \\ $pres:1.75$}  &  \makecell[\bob{}]{discretized to boolean \\for quality questions}   \\ [0.5cm] \midrule

\textbf{\underline{Expert climate}}&  &   &   &    \\ [0.3cm]
\formatvar{tri}{river overflow} & \makecell[\formatres{}]{low, medium,\\ high, none} & \makecell[\formatres{}]{none: $94.94\%$ \\med: $2.44\%$}  &  \makecell[\formatres{}]{none: $90.23\%$ \\med: $4.56\%$} &  -  \\ [0.3cm]

\formatvar{tri}{run off} & \makecell[\formatres{}]{low, medium,\\ high, none} & \makecell[\formatres{}]{none: $99.98\%$ \\med: $0.01\%$}  &  \makecell[\formatres{}]{none: $99.95\%$ \\med: $0.03\%$} &  -  \\ [0.3cm]

\formatvar{ppri}{all but submersion} & 7 classes & \makecell[\formatres{}]{none: $98.08\%$ \\low: $0.75\%$}  &  \makecell[\formatres{}]{none: $93.11\%$ \\high: $2.62\%$} &  \makecell[\bob{}]{classes from very  \\high to none}  \\ [0.3cm]

\formatvar{hydro\_zone}{catchment zones} & $24$ zones & - & - &  - \\ [0.3cm]

\formatvar{clim\_region}{climate regions} & $10$ zones & \makecell[\formatres{}]{\small{plains oc.: $31\%$} \\\small{altered oc.: $18\%$}} & \makecell[\formatres{}]{\small{plains oc.: $36\%$} \\\small{medit.: $14\%$}} &  \makecell[\bob{}]{oc: oceanic \\medit: mediterranean}\\ [0.3cm]

\formatvar{nb\_catnat}{number of accepted catnat}& $[0,40]$  & $5.6~~(4.95)$   & $6.65~~(4.47)$ & \makecell[\bob{}]{used to define historically\\ exposed zones} \\ [0.5cm] \midrule

\textbf{\underline{Rainfall}}&  &   &   &    \\ [0.3cm]
\formatvar{tail\_weight\_cluster}{areas by tail shape} &   & $53.08~~(27.79)$  & $69.09~~(27.58)$  & built with ERA5-land data   \\ 
\formatvar{MILRE}{most intense local rainfall event} & $[14.24,100]$ & $95.53~~(9.65)$ & - & cost data version   \\  
\formatvar{ann\_MILRE}{most intense local rainfall event} & $[0,100]$ & $68.66~~(25.37)$ & $82.64~~(20.22)$  & annual occurrence version   \\ [0.5cm] \midrule

\textbf{\underline{Building and surrounding}}&  &   &   &    \\ [0.5cm]

\formatvar{living\_surface}{m$^2$ living house surface} & $[2.41,2443.52]$  & $112.54~~(56.77)$  &  $118.13~~(59.26)$ & data completed using ML \\ [0.3cm]

\formatvar{house\_value}{value of a m$^2$ of the house}& $[200,28544]$  & $2306.66~~(1237.34)$  &  $2593.38~~(1439.02)$ & data completed using ML \\ [0.5cm]

\formatvar{construction\_period}{construction period}& $[0,2022]$ & \makecell[\formatres{}]{$[0,1915]:26.28\%$ \\$[1949,1968]:11.61\%$} & \makecell[\formatres{}]{$[0,1915]:26.26\%$ \\$[1949,1968]:11.88\%$} & built with text processing \\ [0.5cm] 

\formatvar{nb\_floors}{number of floors}& $[0,4]$ & $0.94~~(0.69)$ & $0.90~~(0.63)$ & data completed using ML    \\ [0.5cm]

\formatvar{outbuilding\_surface}{eventual outbuilding surface}& $[0,1518.03]$  & $9.87~~(29.14)$  & $10.06~~(23.69)$  & built with computer vision\\ [0.5cm]
 \formatvar{wall\_material}{principal wall material}& \makecell[\formatres{}]{\small{agglomere, brick,} \\\small{concrete, etc.}} & \makecell[\formatres{}]{aggl.: $19.53\%$ \\ brick: $18.84\%$}  & \makecell[\formatres{}]{brick: $19.16\%$ \\ aggl.: $16.87\%$}   & built with computer vision \\ [0.5cm]

\formatvar{terrain\_maxslope\_50m}{50m buffer around the building}& $[0.09,307.68]$  & $7.76~~(7.64)$  &  $8.92~~(9.06)$ & \makecell[\bob{}]{maximum slope between \\building and surroundings} \\ [0.5cm]

\formatvar{nb\_building\_50m}{number of building within 50m}& $[0,164]$  & $19.73~~(15.24)$   & $16.83~~(13.38)$ & built with computer vision and gis   \\ [0.5cm]

\formatvar{pres\_adjoining}{presence of adjoining building }&  {pres, abs} & \makecell[\formatres{}]{abs: $60.54\%$ \\ pres: $39.46$}  & \makecell[\formatres{}]{abs: $67.43\%$ \\ pres: $32.57$} & -   \\ [0.5cm]

\formatvar{distance\_watercourse}{distance closest watercourse}&  $[0.0 21909.2]$ & $464.83~~(632.67)$  & $338.88~~(453.32)$  & \makecell[\bob{}]{meters, euclidean distance and\\ all watercourse considered}\\ [0.5cm]

\formatvar{altitude\_diffwatercourse}{altitude difference}& $[-2196.2,525.8]$ & $-12.92~~(36.29)$  & $-9.78~~(29.31)$ & \makecell[\bob{}]{water course bed considered \\for closest point altitude}  \\ [0.5cm]

\formatvar{soil\_type}{predominant soil type} & \makecell[\formatres{}]{\small{sand, clay, marl,} \\\small{limestone etc.}}  & \makecell[\formatres{}]{\small{lst\_marl\_plr: $25.23\%$} \\ \small{sand: $24.92\%$}}  & \makecell[\formatres{}]{\small{lst\_marl\_plr: $31.53\%$} \\ \small{sand: $26.96\%$}} &  unique or combination of taken value   \\ [0.5cm]

\formatvar{impervious\_surface}{percentage around the building}& $[0,1]$ & $0.72~~(0.29)$  &  $0.66~~( 0.31)$  &  in a 200m grid around the building\\ [0.3cm]

\formatvar{length\_partywall}{total wall length}& $[0,661.51]$ & $5.14~~( 8.5 )$  &  $3.98~~(7.42)$  &  calculated using all adjoining buildings\\ \bottomrule

\end{tabular}

}
\caption{Overview of the variables considered in this study. ML for machine learning.}
\label{tab:var_before}
\end{table}

\restoregeometry

\end{document}